\documentclass[a4paper, 11pt]{article}

\usepackage{graphicx}
\usepackage{amsfonts}
\usepackage{amsmath}
\usepackage{amsbsy}
\usepackage{amssymb}
\usepackage{subfigure}
\usepackage{datetime}
\usepackage{geometry}
\usepackage{setspace} 
\usepackage{bm}
\usepackage{multicol}
\usepackage{multirow}
\usepackage{pifont}
\usepackage{color}
\usepackage{framed}

\begin{document}

%%%% Article title to be placed here
\title{An overview of flux braiding experiments}

\author{A.L. Wilmot-Smith}
\date{\small{Division of Mathematics, University of Dundee, Dundee, DD1 4HN.}}

\maketitle

%%%%%%%%% Insert author address here
%\address{Division of Mathematics, Fulton Building, University of Dundee, Dundee, DD1 4HN.}

%%%% Subject entries to be placed here %%%%
%\subject{Solar Physics, Plasmas, Magnetohydrodynamics.}

%%%% Keyword entries to be placed here %%%%
%\keywords{The Sun: Corona, Magnetic Fields, Magnetic Reconnection.}

%%%% Abstract text to be placed here %%%%%%%%%%%%
\begin{abstract}

Parker has hypothesised that, in a perfectly ideal environment, complex photospheric motions 
acting on a continuous magnetic field will result in the formation of tangential discontinuities 
corresponding to singular currents. 
We review direct numerical simulations of the problem and find the evidence points to
a tendency for thin but finite thickness current layers to form, with thickness exponentially decreasing  in time.  
Given a finite resistivity these layers will eventually become important and cause the dynamical process of energy 
release. Accordingly, a body of work focusses on evolution under continual boundary driving. The 
coronal volume evolves into a highly dynamic but statistically steady state where quantities have a
temporally and spatially intermittent nature  and where the Poynting flux and dissipation
are decoupled on short timescales.  
Although magnetic braiding is found to be a promising coronal heating mechanism 
much work remains to determine its true viability.
Some suggestions for future study are offered.

\end{abstract}
%%%%%%%%%%%%%%%%%%%%%%%%%%%

%%%%%%%%%% Insert the texts which can accomdate on firstpage in the tag "fmtext" %%%%%

\section{Introduction}

The notion that solar coronal loops are heated as an end result of magnetic braiding dates back to the 1970s and
Parker's notion of topological dissipation \cite{Parker72, Parker88, Parker94}.
The hypothesis is built on the foundation that the solar corona can be modelled as a largely force-free environment, 
with the Lorentz forces in force balance. However, loops themselves are subject to photospheric motions at their footpoint 
and hence slow footpoint motions will lead to the quasi-static 
evolution of loops through sequences of force-free equilibria.  Complex footpoint motions applied to the base of a coronal
loop will twist and tangle the magnetic field which must then relax to a force-free state. 
Given the extremely high Lundquist numbers of the corona the relaxation will 
 be ideal and so will preserve exactly the magnetic field topology.

\begin{figure}[hbtp]
   \centering
   \includegraphics[width=0.32\textwidth]{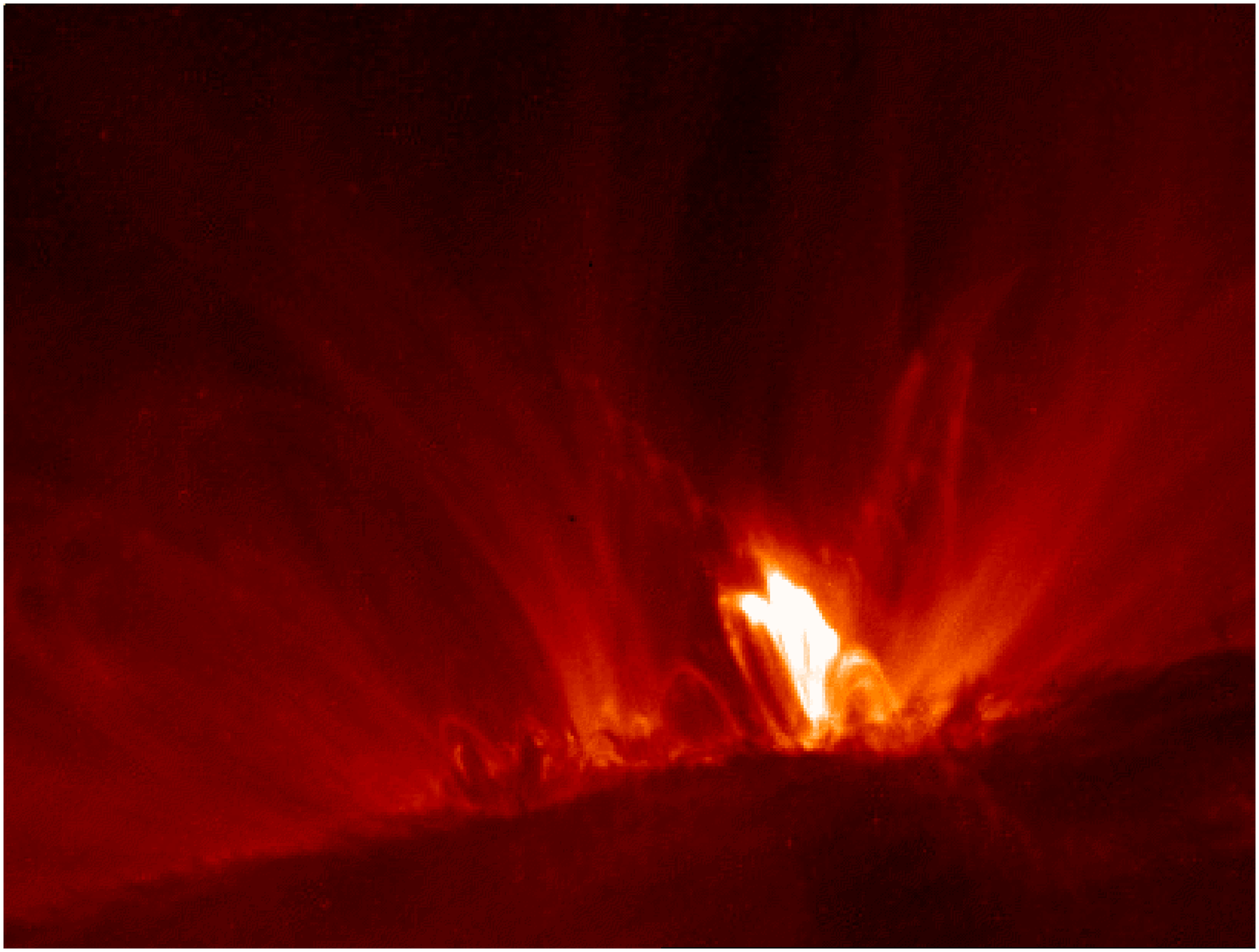} 
      \includegraphics[width=0.32\textwidth]{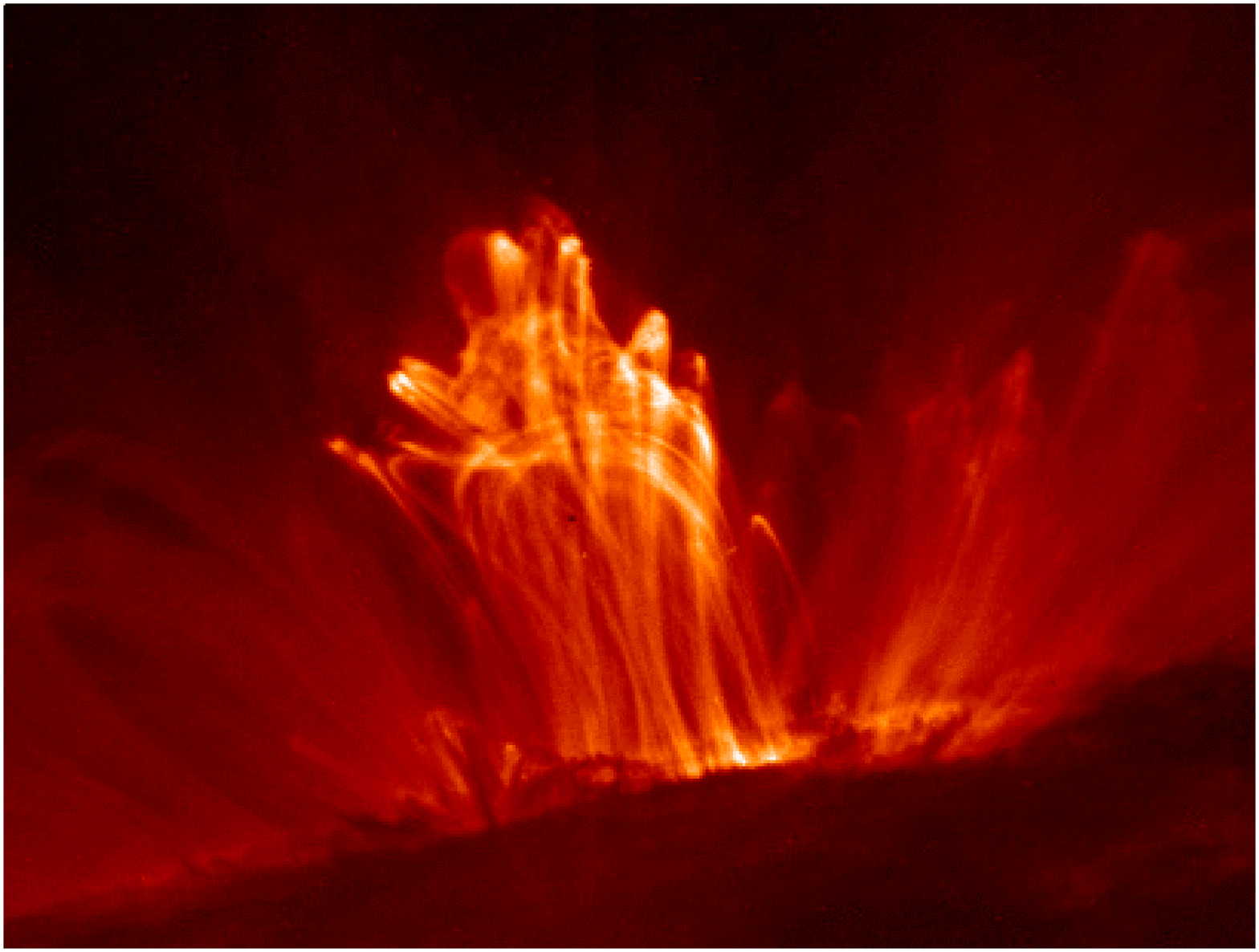} 
   \includegraphics[width=0.32\textwidth]{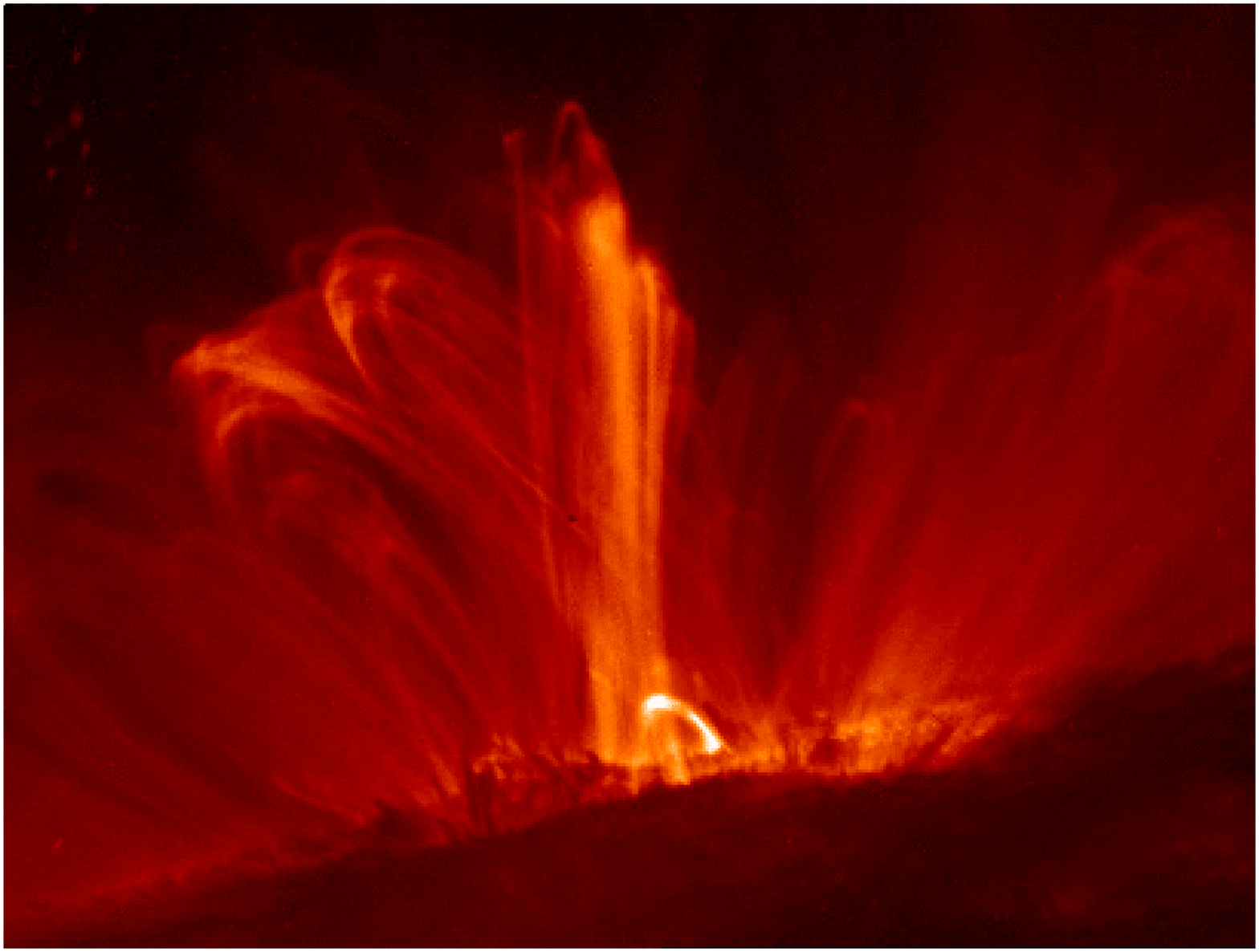} 
   \caption{Evolution of an M-class solar flare observed by the \textit{Transition Region and Coronal Explorer}
   at 06:34UT, 08:13UT and 09:03UT on 14th September 2000 in the 171\AA \ passband.  An apparently 
   tangled magnetic field structure relaxes to a simpler configuration after the flare.}
   \label{fig:trace}
\end{figure}

The question is then whether a magnetic field of arbitrary topology can relax ideally to a smooth force-free equilibrium.
Parker hypothesised that the space of force-free fields is restricted, so that smooth equilibria will not generally exist
and, instead, the magnetic field will develop tangential discontinuities corresponding to current sheets \cite{Parker72}.  
Of course in a real corona as soon as sufficiently small length scales develop then diffusion will locally become appreciable and
a change in the magnetic topology can occur, releasing energy.  This release of magnetic energy, built up through braiding,
could, under the Parker hypothesis, explain the observed high temperatures of coronal loops.

Observing braided magnetic fields in the corona has so far been fraught with difficulty.  Although a very few observations 
(such as those of Figure~\ref{fig:trace}) suggest large-scale magnetic fields can have braided configurations, most 
observations of coronal loops show an apparently well-combed structure.  The latest high-resolution images
from Hi-C are more suggestive of braids \cite{Cirtain}, although how the apparently crossing structures relate to the underlying
magnetic field is not immediately clear \cite{Cargill}.  A non-linear force-free extrapolation of 
the region \cite{Thal} does suggest a complex structure.  Although the extrapolated magnetic field outlines 
the crossings seen in the observations, it is also constructed from lower-resolution magnetograms.
 Furthermore, 
the flux tube determined is split along its length which can only be the result of at least one coronal null point.
A null point in such a coronal volume is expected from our current understanding of the structure of 
coronal fields:  they are known to be permeated by nulls, separatrix surfaces and separators \cite{Priest14}.  
A review in this volume \cite{Parnell}
 discusses the significance of heating at these features, which was already acknowledged by Parker:  
\textit{ ``Insofar as the field is concentrated into separate individual magnetic fibrils at the photosphere, each individual fibril moves independently of its neighbors, producing tangential discontinuities (current sheets) ... There is, however, a more basic effect, viz., a continuous mapping of the footpoints spontaneously produces tangential discontinuities''} \cite{Parker88}.
Accordingly, braiding as discussed in this review refers specifically to the elementary effect in individual elemental 
loops.  These loops are not resolved by current instruments and, given their aspect ratio, fine scale braiding within 
a loop would likely appear smoothed out.

The notion of footpoint motions acting on the complex coronal field with all of its topological features has been formalised
into the theory of coronal tectonics \cite{Priest}.  Simulations addressing this relevant scenario are so far rare, with 
just a very few examining the basic effect  \cite{Mellor, IDM06a, IDM06b, WS07}.
  One important series of articles implicitly fall into this framework \cite{GN02, PGN04, GN05, GN05b}.
Here a potential magnetic field is extrapolated into the corona from a smoothed active region magnetogram 
and a convection-like driving velocity applied to its simulated photospheric boundary.
Although the resolution of these large-scale simulations is not sufficient to provide an understanding of exactly where
or how the dissipation is occurring, the broad comparisons to observed loops are encouraging.

An understanding of the viability of the Parker mechanism for coronal heating requires an two-pronged attack both by 
theory and by direct numerical simulation, while this review discusses only the latter approach.
As we will detail in Section~\ref{sec:overview}, numerical evidence does not, at least to date, 
demonstrate the ubiquitous formation of 
singular current sheets.  Nevertheless, simulations all agree that small scales in the current do rapidly develop 
under generic braiding conditions and these will eventually become small enough to 
initiate dissipation.  Hence a second main question to be tackled by braiding simulations arises:  
given a finite resistivity,  how does the coronal magnetic field evolve when it is continuously driven by photospheric
motions?
A discussion of the main simulations tackling these various questions is given in Section~\ref{sec:overview},
presented in four broad classes according to simulation setup.  Conclusions are briefly drawn in 
Section~\ref{sec:conclusions} together with an outline of some suggestions for future work.

\section{Braiding Simulations}
\label{sec:overview}

Numerical simulations of flux braiding naturally divide into four categories.  These
 are not all distinct, with some simulations falling into more than one category.  Nevertheless the division is
helpful to describe the present state of knowledge.  In brief, the groups of simulations are: 
\begin{itemize}
\item[(a)] \textit{Sequences of shears}.  A loop is subjected to a sequence of simple shearing motions
on the boundary. \\[-10pt]
\item[(b)] \textit{Continually driven systems}.  A loop is subjected to boundary motions, generally of rotational form, 
for an extended period of time. \\[-10pt]
\item[(c)] \textit{Formation of discontinuities}.  The question of whether or not tangential discontinuities form in a coronal
volume subjected to boundary motion is examined by simulation. \\[-10pt]
\item[(d)] \textit{Initially braided fields}.  The coronal volume is not braided self-consistently via boundary motions
but a braided magnetic field is taken as an initial condition for a simulation.
\end{itemize}

Each of the simulations discussed in this review have a broadly similar experimental setup.
To model the coronal loop the magnetic field is straightened out to lie between two parallel plates.  
Hence a Cartesian geometry is employed and in the numerical box both the upper and lower boundaries 
represent the photosphere.  A velocity field imposed on one or both boundaries represents a photospheric flow.  
The flow is two--dimensional and flux emergence (or cancellation) is not considered.
We  detail results found for simulations in each of the classes given above in the following sections, beginning 
with the earliest form of simulation, that of boundary shear.

\subsection{Sequences of boundary shears}
\label{subsec:shear}

\begin{table}[!h]
\caption{Outline of braiding simulation setups investigating
evolution when shearing motions are applied to loops.}
\begin{tabular}{ p{2cm} p{1.7cm}   p{1.43cm}  p{1.1cm}    p{5.8cm}   }
\hline
Article & Evolution Method & Grid size & Box size & Driver properties  \\ \hline

van Ballegooijen \cite{vB88a,vB88b}
 & magnetic energy minimisation & N/A & $1^{3}$ &
 Upper boundary only: sequence of 5 low amplitude perpendicular shears
with random phase angle\\[2pt]

Miki{\'c}, Schnack, van Hoven \cite{MSvH} & Simplified 3D ideal MHD & 
$64^{3}$ & $1^{3}$ & Lower boundary only:  12 low amplitude  perpendicular flows based on 
\cite{vB88a, vB88b}\\[2pt]

Longbottom \textit{et al.} \cite{L98} & magneto-frictional & up to $65^{3}$ & $1^{3}$ & 
Fixed high amplitude initial shear, second perpendicular shear of varying amplitude. \\[2pt]

Galsgaard \& Nordlund \cite{GN96} & 3D resistive MHD & up to $136^{3}$ &
$1^{3}$ or $1^{2} \times 10$ & Both boundaries driven, random amplitudes, phases
and durations.  \\[2pt]

Bowness \textit{et al.} \cite{B13} & 3D ideal/ resistive MHD &
$512^{3}$ & $1^{2} \times 4/3$ &
Two perpendicular shears (second switches off in one study, continues in another). \\

\hline
\end{tabular}
\label{tab:sheardriv}
\end{table}%%%End of the table

\begin{figure}[bthp]
   \centering
   \includegraphics[width=0.3\textwidth]{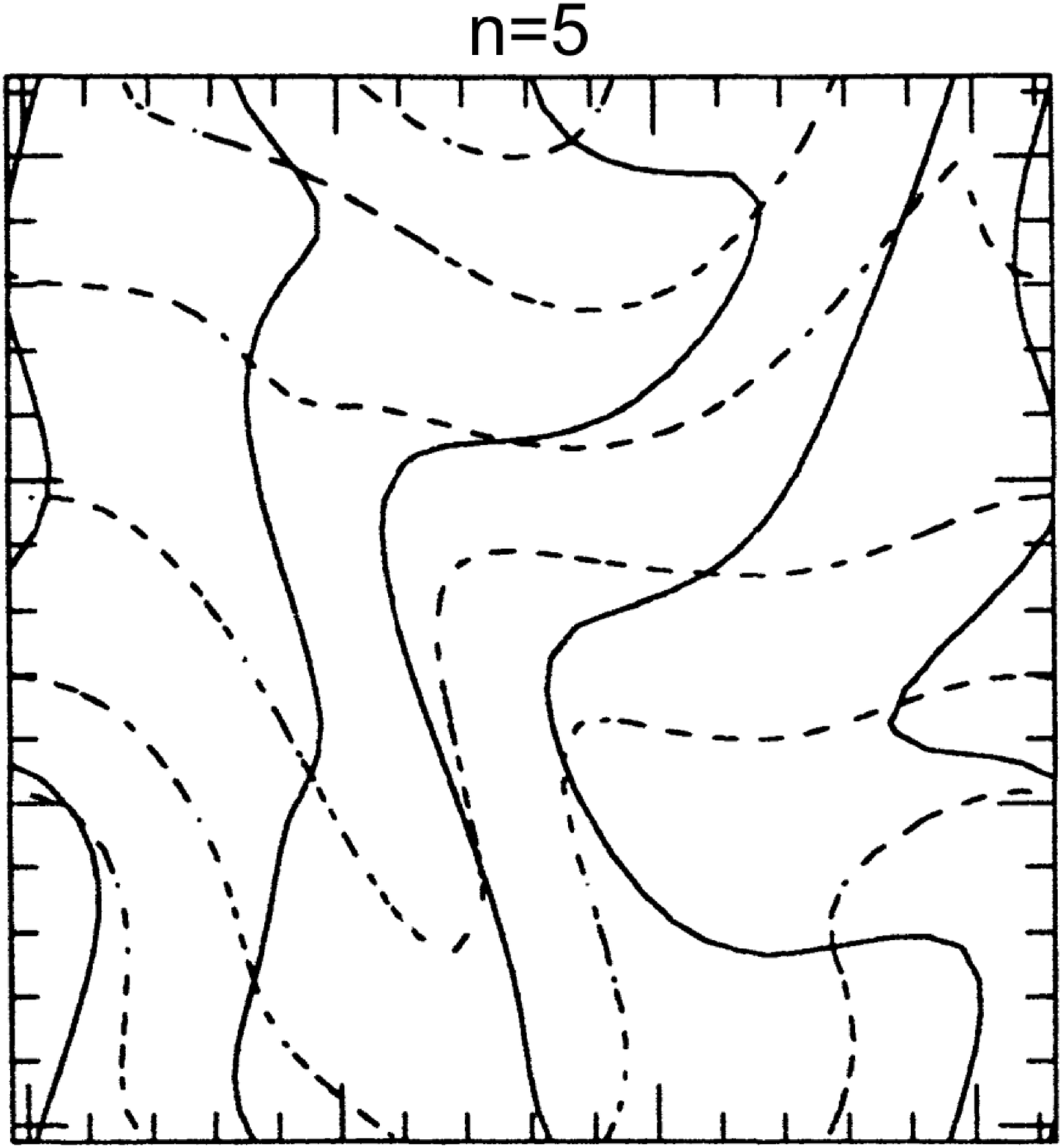} 
      \includegraphics[width=0.3\textwidth]{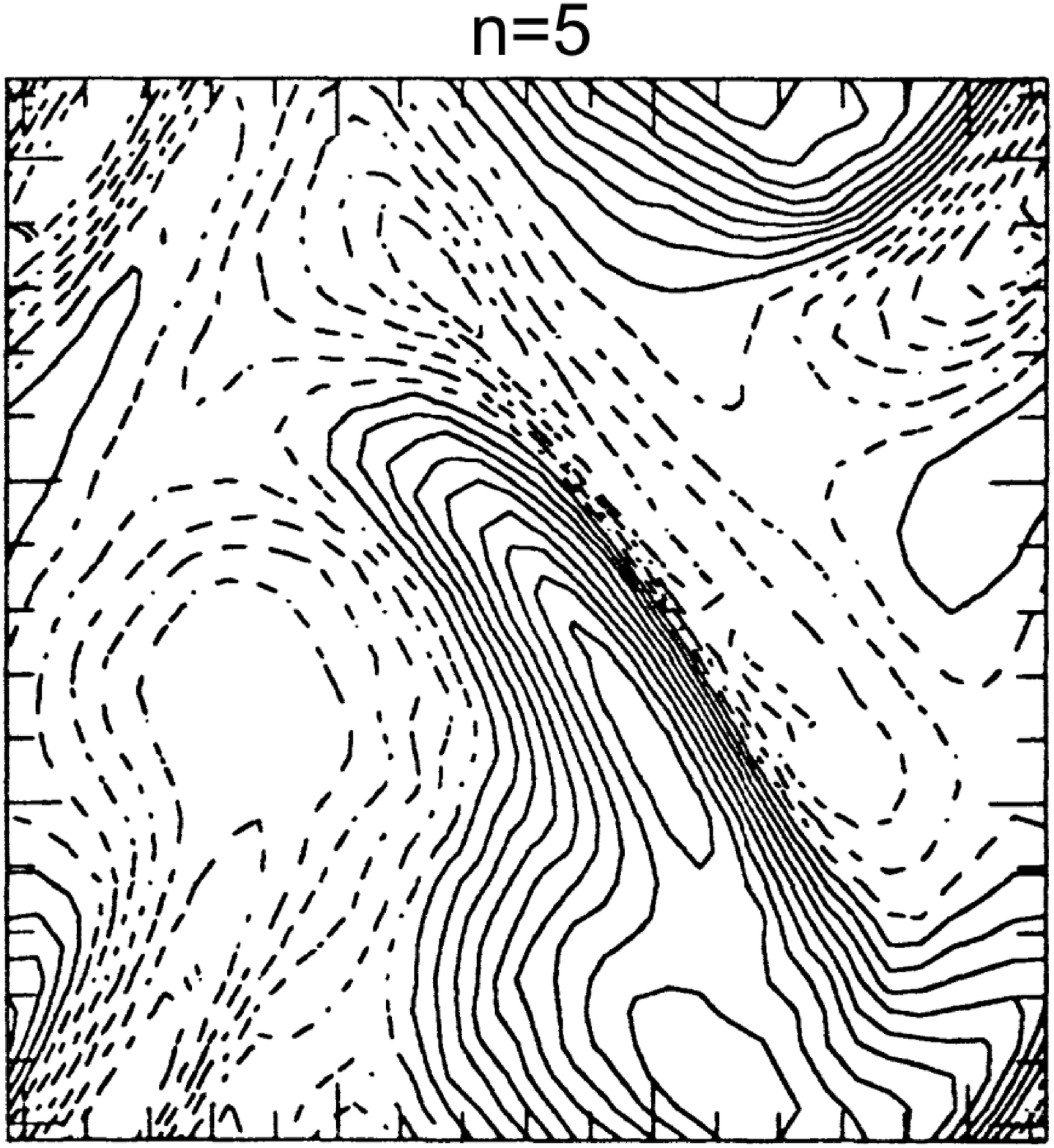} 
   \caption{Illustrating the nature of the force-free equilibria attained by van Ballegooijen \cite{vB88b}
   following 5 boundary shearing motions applied to an initially homogeneous field.
Displacement of fluid elements on the upper boundary (\textit{left}) and vertical component of the
electric current near the lower boundary (\textit{right}). Image adapted from  \cite{vB88b}. }
   \label{fig:vB88}
\end{figure}

Table~\ref{tab:sheardriv} provides a short summary of simulation setups 
where initially uniform magnetic fields are subjected to shearing motions on their boundaries.  
Here, and throughout this section, the term shearing flow is used to describe one running on opposite directions 
on either side  of a neutral flow line.
Van Ballegooijen \cite{vB88a,vB88b} was the first to describe such a simulation, applying a shearing 
flow on the lower boundary of the domain in an alternating sequence of perpendicular directions
 (${\bf e}_{y}$, ${\bf e}_{x}$, ${\bf e}_{y}$, etc.) with differing, randomly chosen, phase angles.  
 The flow amplitude and duration were
each chosen such that the maximum displacement of fluid elements on the boundary is around $16\%$ of the domain size.
 The mapping of magnetic field lines from lower to upper boundary is shown to develop fine scales in the 
evolution,   with the scales decreasing exponentially with number of shear events \cite{vB88a}.  
Although this work was developed before
  the concept of the quasi-separatrix layer (QSL) was developed \cite{THD02}, the idea is essentially the same:  QSLs
  in the domain have a thickness exponentially decreasing in time.
  Fluid displacements on the upper boundary after 5 shears are shown in the left-hand image of Figure~\ref{fig:vB88}.
%  (the gradients of these mappings having small scales).
  
  The corresponding volume force-free magnetic field was obtained following each boundary displacement
  using an energy minimisation iterative procedure on a $32^{3}$ numerical grid \cite{vB88b}.
Using this approach a total of five shears were successfully applied before the method itself failed.
While it is not \textit{a priori} clear that the existence of small scales in the magnetic field line mapping also
forces the magnetic field and current to have the same small scales, this does turn out to be the case in 
the example presented \cite{vB88b}.  Hence the length scales of the current also exponentially decrease
with number of shears. The current structure after 5 shears is shown in 
the right-hand image of Figure~\ref{fig:vB88}.

Miki{\'c}, Schnack \& van Hoven \cite{MSvH} took a similar approach, again applying a shear sequence to the
lower boundary of a unit cube containing a magnetic field.  Their numerical scheme, a simplification of the
ideal MHD equations that neglects the inertial term and employs a high viscosity, allows for a successive 
sequence of 12 force-free relaxations on a $64^{3}$ grid to be calculated.  The main findings of 
van Ballegooijen \cite{vB88a, vB88b} were confirmed, with the cascade to smaller scales giving an exponential growth
in current density but with a smooth force-free equilibrium achieved in each step.
A consequence of the exponential decrease in current layer thickness with shear is that a tangential discontinuity
will only be reached after an infinite time.  However, in practice current densities 
will be strong enough to become dynamically important in a plasma of finite resistivity.

While the  shears applied in  \cite{vB88a, vB88b, MSvH} all have a low
 amplitude ($< 20\%$ of the domain size),  Longbottom \textit{et al.} \cite{L98} considered the question 
of how the shear amplitude itself (rather than number of shears) might affect the field evolution.
The authors employed an ideal magnetofrictional relaxation method \cite{CS86}, and applied
shearing motions on both the upper and lower boundaries of the domain.  The initially uniform magnetic 
field on the unit cube was first subjected to a shear of amplitude 0.8 (as a fraction of the domain size) 
 and then relaxed to a force-free state that was found to be smooth.
This state was taken as an initial condition for a parameter study in which a perpendicular shear of various 
strengths was applied before relaxation.   
The  amplitude of the second shear is shown to be crucial in determining the nature of the final force-free state.
In all cases a twisted current structure is found running through the domain.
The basic structure of the twisted current layer was confirmed by Bowness \textit{et al.} \cite{B13}
who imposed analytically an initial shear and then used an ideal 3D MHD code at $512^{3}$ to evolve the magnetic field 
through a second, perpendicular, shear event.  Taking a second shear strength of 0.5 a
twisted current layer forms (Figure 11(a) and Figure~14 of \cite{B13}) running through the domain, agreeing with \cite{L98}.

The width of the current layer depends on shear strength \cite{L98}:
 for low shear values (below 0.5) the layer is well resolved while for high shear (above 0.6) its behaviour
 is consistent with that of a true tangential discontinuity.  
 That is, the maximum current in the layer increases linearly or faster
 as the grid resolution is increased (power law growth is expected for a true current sheet \cite{AS01}).
The simulation of Longbottom \textit{et al.} \cite{L98} marks the first (and essentially only)
demonstration of a case that is consistent with Parker scenario of topological dissipation \cite{Parker72}.  
With the resolution available at that time (a Lagrangian grid of maximum size $65^{3}$) 
and known inaccuracies in the scheme for high grid deformation \cite{P09}, the possibility
remains that the growth of maximum current with resolution in 
the high shear cases could level off at higher resolutions.
%That is, the behaviour suggestive of a discontinuity is also consistent with a smooth but small scale 
%current layer that only becomes resolved at higher grid resolution.
An increase in grid size has only very recently become accessible \cite{C14} and this possibility is, as-yet, untested.

Under either of these aforementioned scenarios (an exponential decrease in current thickness or the formation
of singular current layers in ideal MHD) in a real physical plasma the resistivity will at some point
become important.  To determine the consequences of a finite resistivity, Galsgaard \& Nordlund \cite{GN96} employed 
a fully resistive 3D MHD simulation to examine behaviour of a coronal loop continually subjected to boundary
shearing motions. The authors detail an extensive set of simulations, common to all of which is the application of 
perpendicular shears of random amplitude, phase and duration to both the upper and lower boundaries of the model loop.
Runs consider the effect of driver speed ($0.02-0.4$ compared with the Alfv{\'e}n speed of 1)
 and duration, loop aspect ratio, and grid resolution ($24^{3}-136^{3}$), 
 with all runs extending for many Alfv{\'e}n loop crossing times.

\begin{figure}[htbp]
   \centering
   \includegraphics[width=0.8\textwidth]{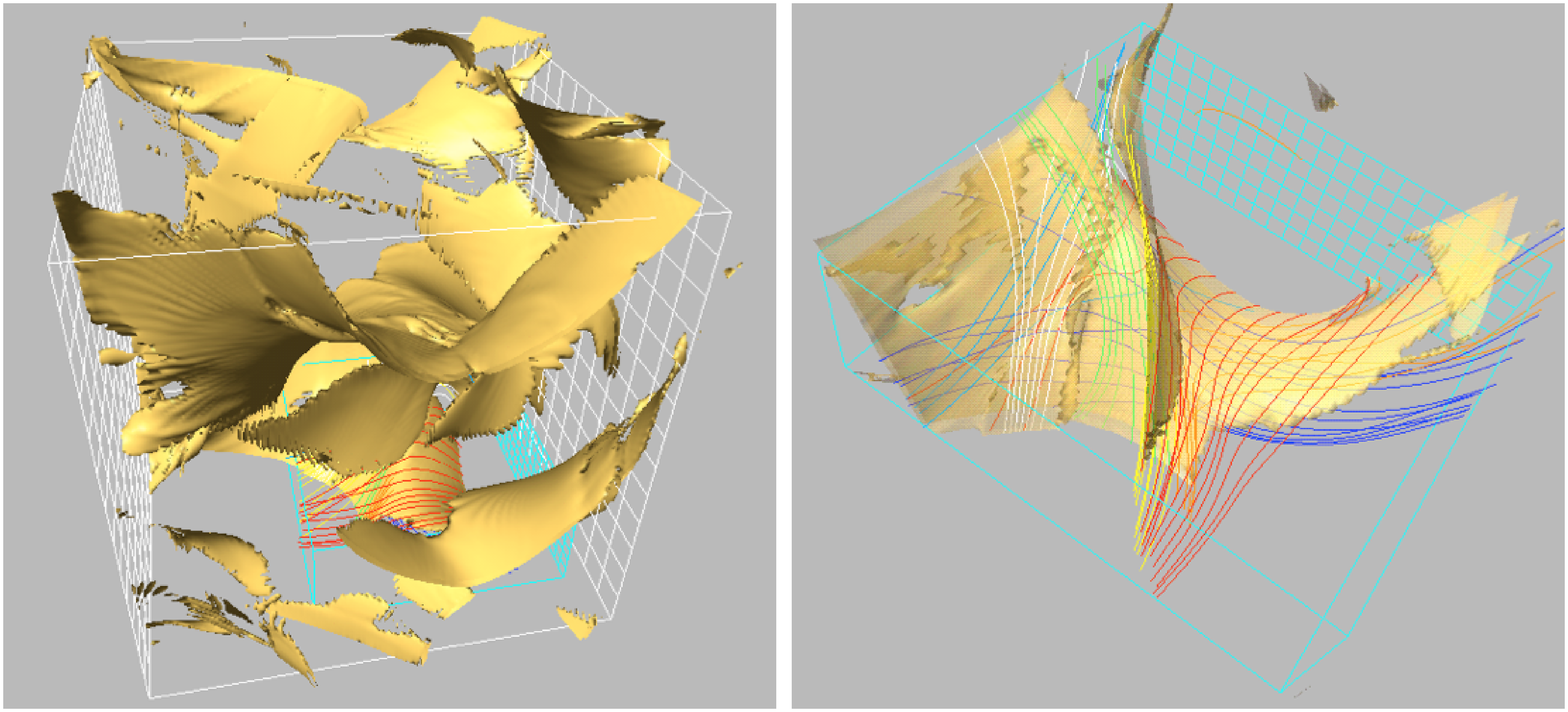} 
   \caption{Instantaneous isosurfaces of current at a representative time in a simulation
  where a loop is continually driven by shearing boundary motions \cite{GN96}.
      The left-hand image shows the full domain (boundary driving surfaces indicated with a grid)
   and the right-hand image a sub-section (the blue box of the left-hand image).
      The current has a fully three dimensional and space-filling structure.  Image reproduced with 
      permission from \cite{GN96}.}
   \label{fig:GN96}
\end{figure}

Some results are common to each of the simulations in \cite{GN96}:
shearing motions cause a rapid growth in the electric current (with maximum strength increasing exponentially
in the early phase), dissipation becomes important and  within two or three shears a statistically steady state 
is reached where quantities (including the maximum current and total magnetic energy) fluctuate about
an average level.
In such steady states the current structure is fragmented with Joule  dissipation taking place over a wide range of 
scales.  Figure~\ref{fig:GN96} shows isosurfaces of current in one particular run. Dissipation has
an increasingly bursty character for the higher box aspect ratio cases.  The magnetic field structure (some example field 
lines are shown in Figure~\ref{fig:GN96}) is generally complex, including reversals in the field component perpendicular
to the driven boundaries.  No significant twist is built up in the system but magnetic energy in excess of potential in 
the steady states varies significantly.  The key factor appears to be the boundary driving velocity:  for the 1:10 aspect ratio
loop the mean energy in excess of potential in the statistically steady states  is just 1.5\% for the slower driver of
at a 0.02 fraction of the Alfv{\'e}n velocity but 45\% for the faster driver at a 0.2 fraction of the Alfv{\'e}n velocity.

The simulation series of Galsgaard \& Nordlund  \cite{GN96} is one in a wider class of simulations
in which a coronal volume is continually driven under a resistive evolution.  Such simulations are
the subject of the next section.

\subsection{Continually driven systems}
\label{subsec:cont}

\begin{table}[thbp]
\caption{Outline of braiding simulation setups investigating
continually driven coronal loops.}%%%Table caption goes here
\begin{tabular}{  p{2cm}  p{2cm}  p{1.55cm}  p{1.1cm}  p{5.6cm} }%%%The number of columns has to be defined here
\hline
Article & MHD evolution type & Grid size & Box size  & Driver properties \\ \hline
        
Longcope \& Sudan \cite{LS93}
 & reduced  & $16^{2} \times10$ to $48^{2} \times 10$ 
&  $1^{3}$ & Time-dependent rotational cells, both boundaries, 20--150  crossing times. \\[5pt] 
   
 Galsgaard \& Nordlund \cite{GN96} & resistive  3D & 
 up to $136^{3}$ & 
  $1^{3}$ and \newline $1^{2} \times 10$ & 
Time dependent shear sequences, both boundaries, typically $\sim 30$
 crossing times.
   \\[5pt] 
    
Hendrix \& van Hoven \cite{HH96} & simplified 3D & 
$32^{2}\times31$ to $256^{2}\times31$
& $(2\pi)^{2} \times 8\pi$ &
Time-dependent rotational cells, both boundaries, 600 crossing times.\\[18pt]   
                        
    Gomez \textit{ et al.} \cite{G00} & reduced & $192^{2}\times32$, $384^{2} \times 32$ & $(2\pi)^{2} \times 10$
   & Stationary rotational cells,  upper boundary, $\sim$ 100  crossing times. \\[5pt] 
                        
Rappazzo \textit{ et al.} \cite{R07, R08, R10, R13}
 & reduced & 
typically $512^{2} \times200$ & $1^{2}\times10$ & 
-- Stationary rotational cells, both boundaries, $\sim$ 600  crossing times \cite{R07, R08}.

--  Comparison cases:  uniform shear \cite{R10}, uniform single localised vortex \cite{R13}.
\\[5pt]   
            
Ng \textit{ et al.} \cite{Ng11} & reduced &$64^{2}\times16$ to $512^{2}
\times64$ & $1^{3}$& Time-dependent rotational cells (as \cite{LS93}),
both boundaries, 10000  crossing times.  \\[5pt] 
    
Dahlburg \textit{ et al.}  \cite{D12} & resistive (with conduction, radiation)
 & $128^{3}$ & $1^{2} \times 5$ & Time-dependent rotational
cells, both boundaries, 700 crossing times. \\[19pt]  
                
Bowness \textit{ et al.} \cite{B13} &  ideal and resistive 3D & $512^{3}$ & $1^{2}\times \frac{4}{3}$ 
&Two perpendicular shears, second shear indefinite or stops, both boundaries, 25  crossing times. \\ 
\hline
\end{tabular}
\label{tab:contdriv}
\end{table}%%%End of the table

An outline of the continually driven simulations discussed here  is given in Table~\ref{tab:contdriv}.
Strong currents rapidly build up in these systems and so in general resistive schemes are required 
to consider the longer-term evolution.  The schemes taken vary but a frequent choice, largely for reasons 
of computational efficiency, is to follow the evolution in a (resistive) reduced MHD (RMHD) scheme 
\cite{Strauss76, Strauss77, Rosenbluth76}.
Under the RMHD assumption an ordering to the system is assumed.
The axial magnetic field component $B_{0} {\bf e}_{z}$ remains constant
while the perpendicular component ${\bf b}$ varies in space and time (its vector potential is evolved)
and is such that $\vert{\bf b}\vert/B_{0} \approx \epsilon \ll 1$.
The velocity field is forced to be incompressible and of order $\epsilon$
and the result is a nonlinear system of two equations that evolve the vorticity and the magnetic vector potential.
The current itself is only in the vertical ${\bf e}_{z}$ direction (but depends on all three coordinates).
There is typically no energy equation in an RMHD evolution so that heat from any dissipation in the system is 
immediately drained away.

\begin{figure}[htbp]
   \centering
   \includegraphics[width=0.24\textwidth]{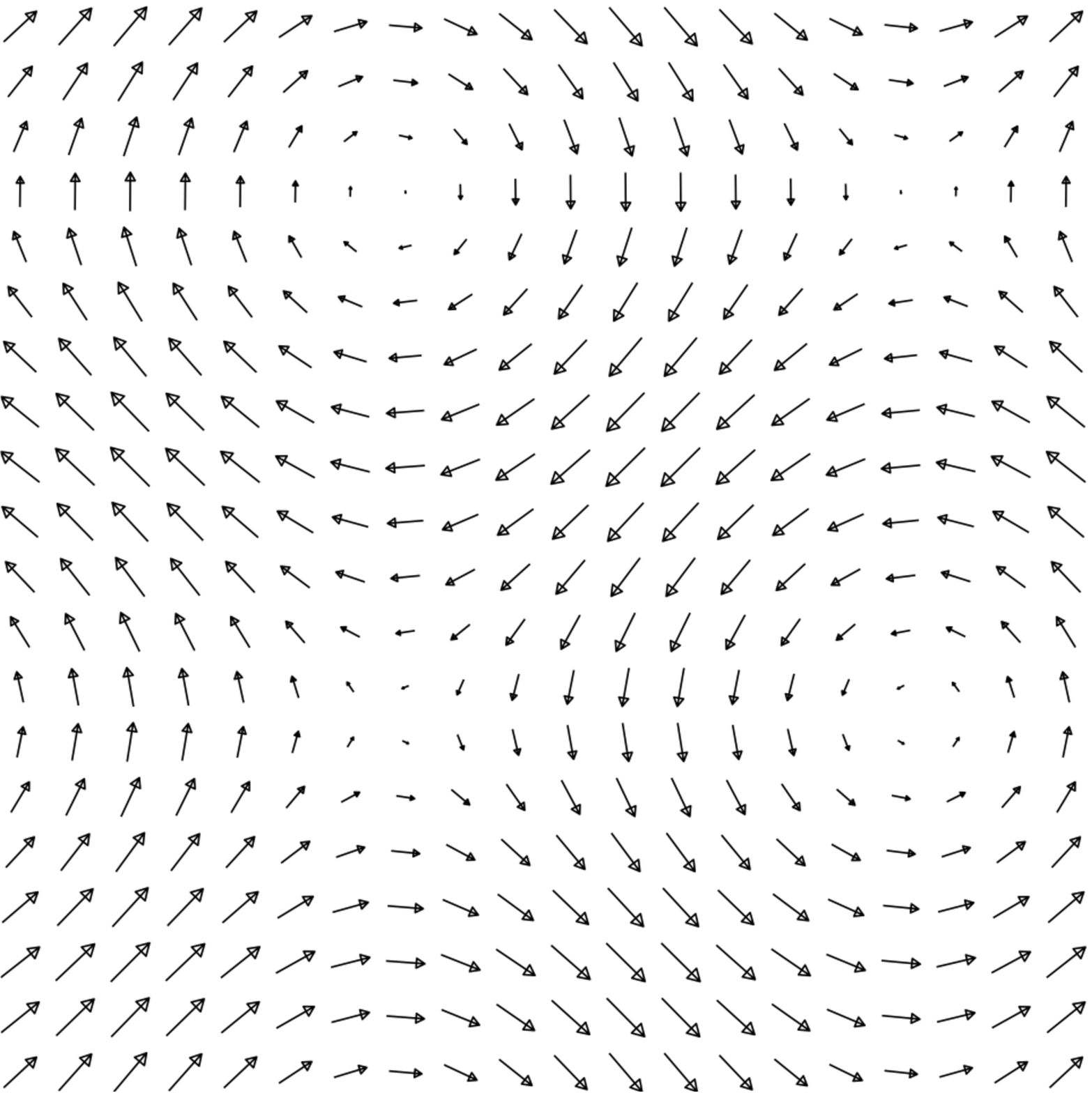} 
      \includegraphics[width=0.24\textwidth]{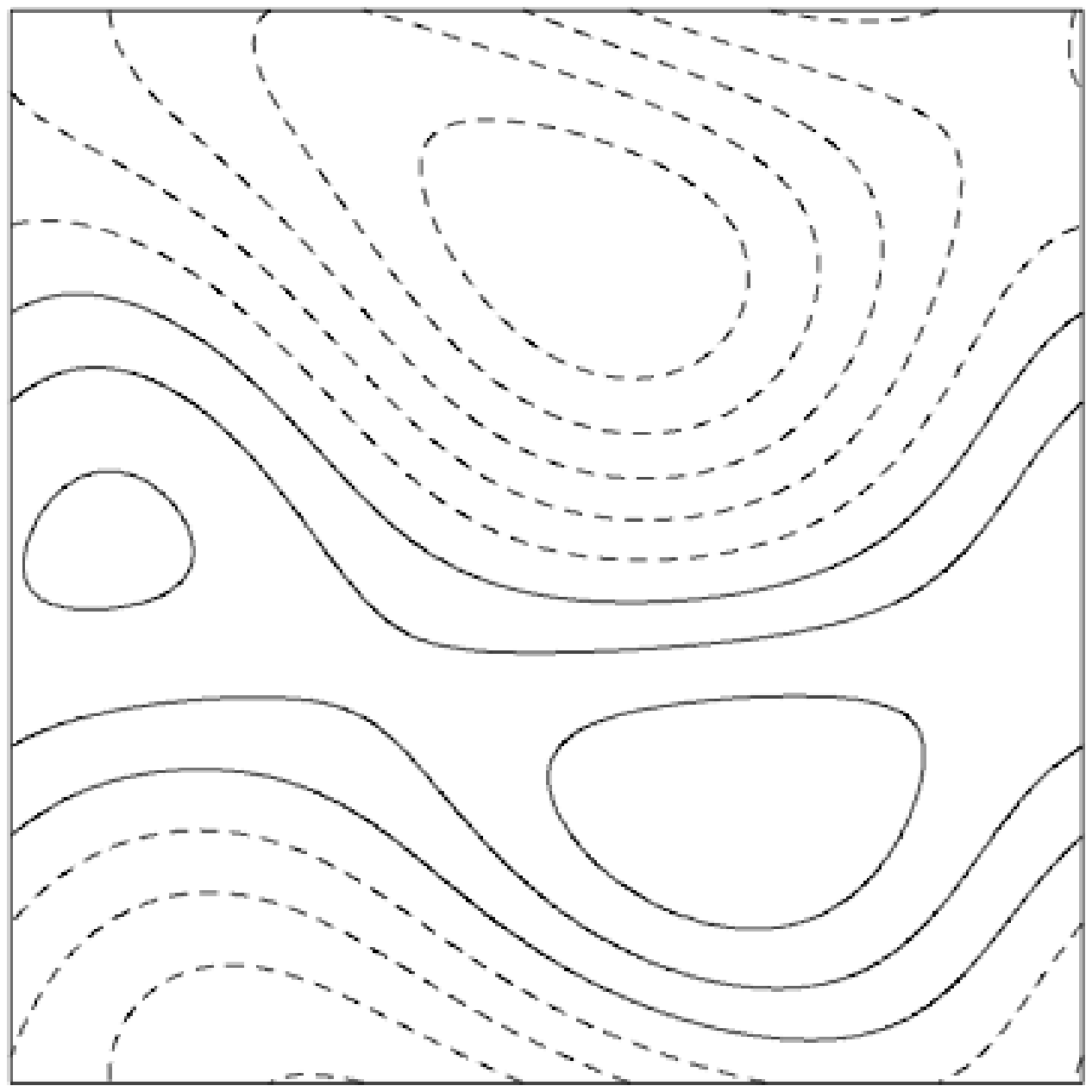} 
      \includegraphics[width=0.24\textwidth]{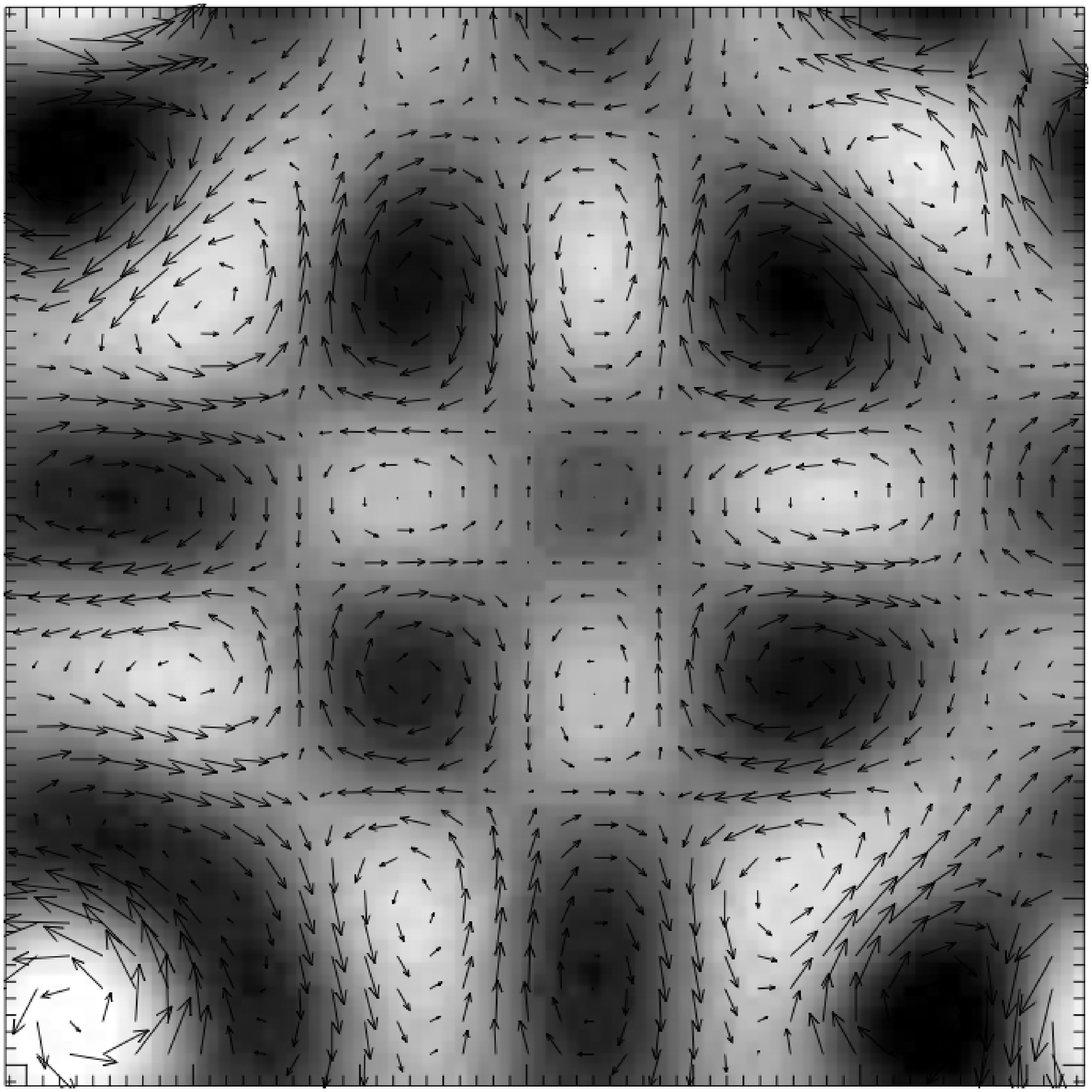} 
      \includegraphics[width=0.24\textwidth]{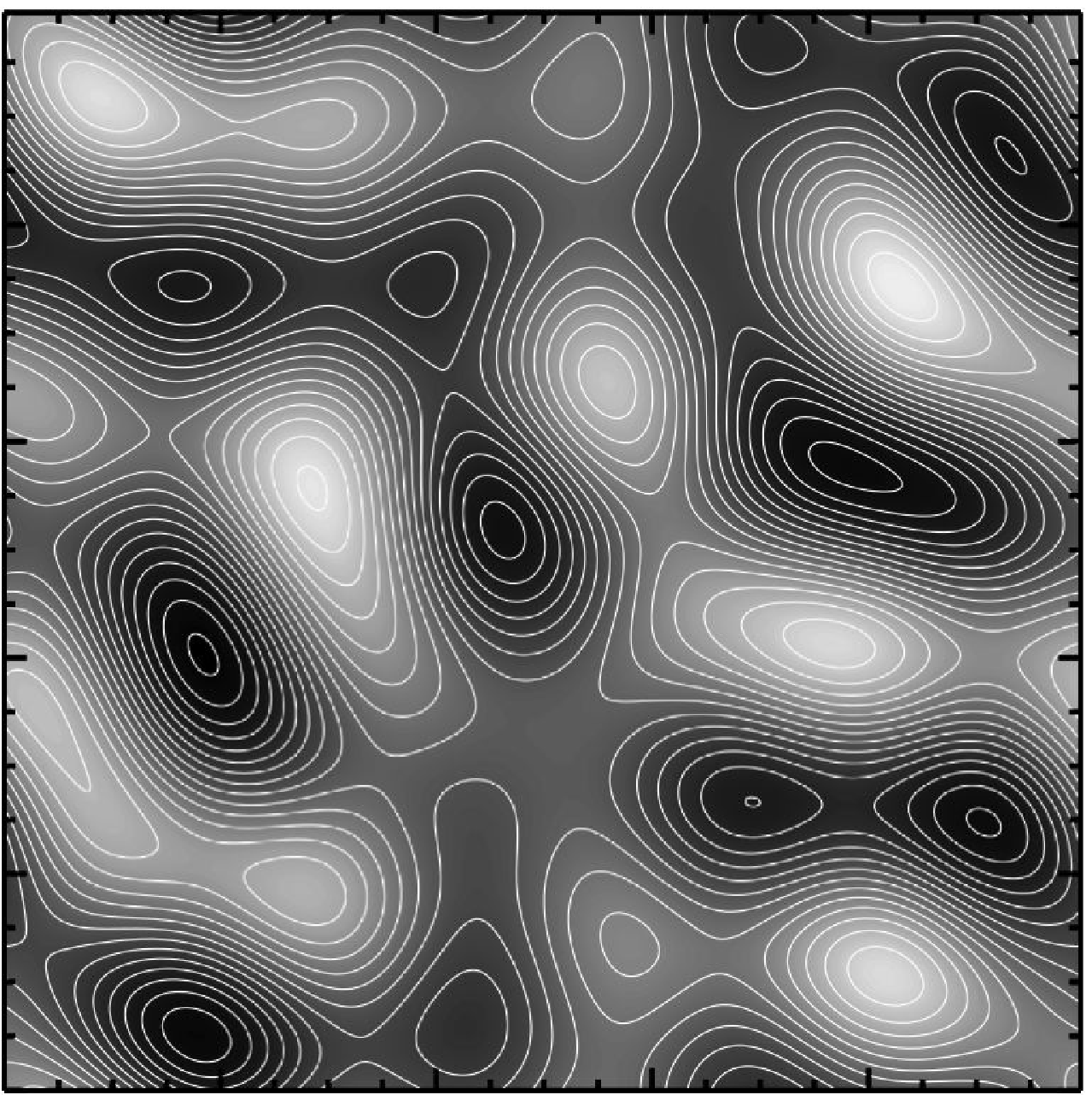} 
   \caption{Streamlines and vector fields at particular times of the boundary driving velocities
   applied to coronal loops in various continually driven simulations. Left \cite{HH96} (particular frame
   from time-dependent simulation); second from left is a frame from the time-dependent 
   driver of \cite{LS93} and \cite{Ng11} (reproduced with permission from \cite{LPhd}), 
   second from right \cite{G00} (stationary driver)
   and right \cite{R07, R08} (stationary driver adapted from \cite{R08}).}
   \label{fig:contdrivers}
\end{figure}

Of the continually driven systems outlined in Table~\ref{tab:contdriv}, two \cite{GN96, B13} are
shearing experiments and have been described in Section~\ref{subsec:shear}.
The remaining works \cite{LS93, HH96, G00, R07, R08, Ng11, D12}
have footpoint motions that are broadly similar, all being modelled
on incompressible projections of convection.  The convection of solar granulation is naturally
compressible and allows for flux emergence and cancellation.  Braiding simulations exclude these
phenomena and so motivate the use of rotational cellular footpoint motions in simulations
(an incompressible model for convection).
Examples of such footpoint motions are illustrated in Figure~\ref{fig:contdrivers}. 
Note that Longcope \& Sudan \cite{LS93} and Ng \textit{et al.} \cite{Ng11} 
both use the same driver  which is time-dependent  and derived from a source of stationary random noise
(illustrated at one particular time in Figure~\ref{fig:contdrivers}, second from left).

In addition to the rotational drivers, Rappazzo \textit{et al.} (\cite{R10, R13}) present two comparison cases.
These are stationary drivers, one with a constant shearing profile \cite{R10} and the other a localised single
vortex motion \cite{R13}.  The motivation for these simple drivers would initially be to produce instabilities (tearing mode
and kink instabilities respectively) rather than as relevant profiles for field line braiding.  
We include these cases here to motivate a discussion of how the driver itself affects evolution.

\begin{figure}[htbp]
   \centering
   \includegraphics[width=0.48\textwidth]{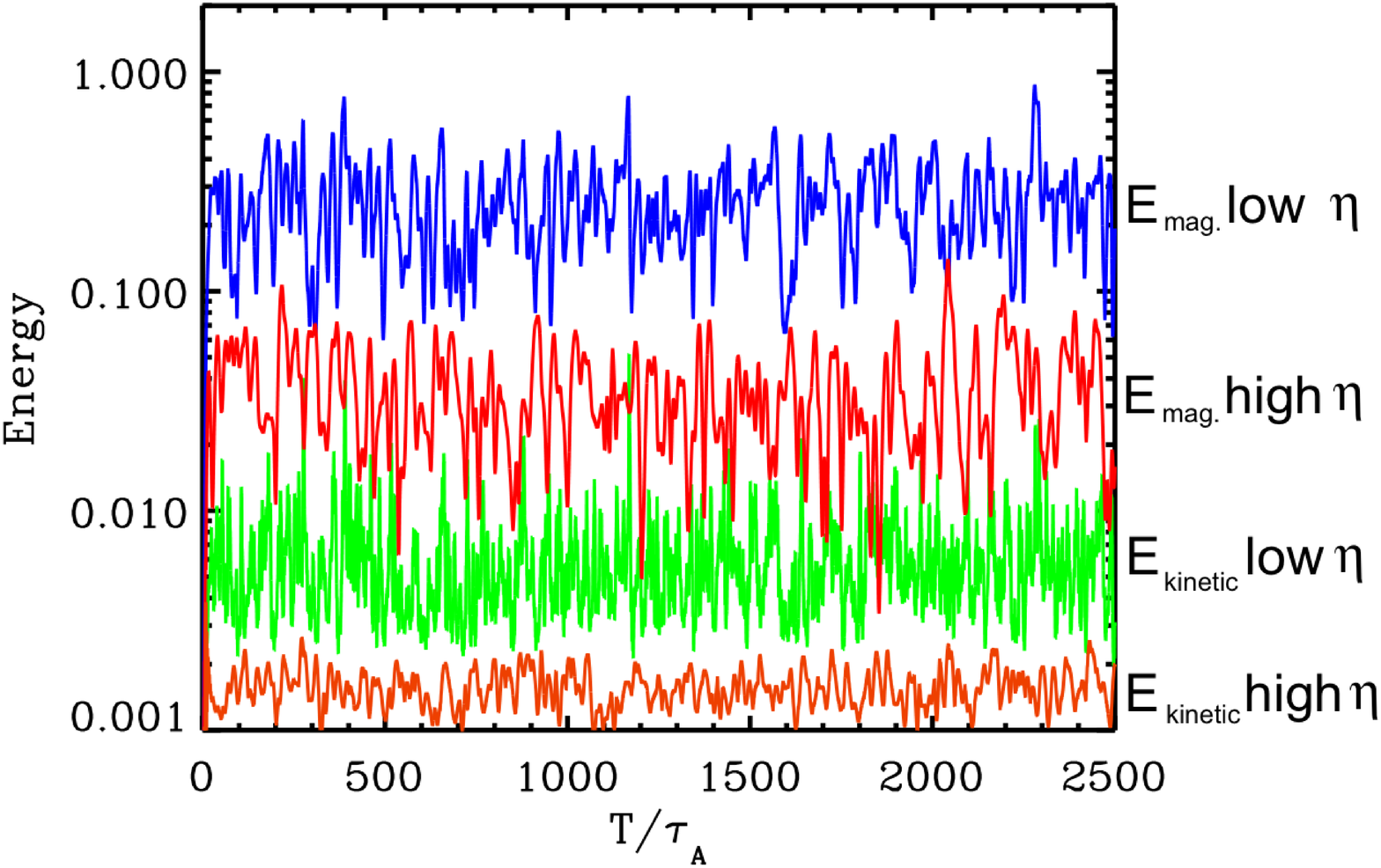}  \ \ \ \ \ 
   \includegraphics[width=0.4\textwidth]{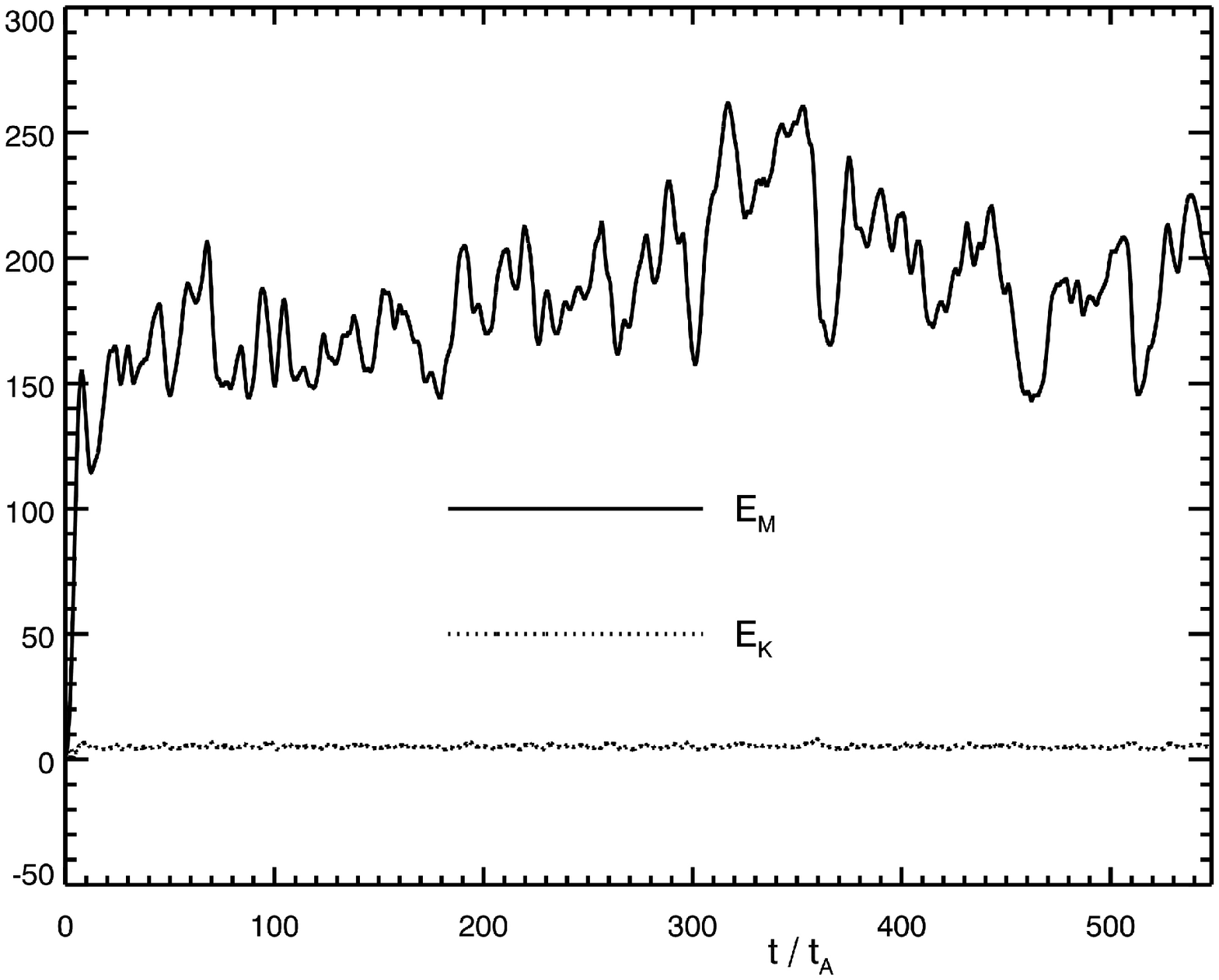} 
   \caption{Variation of total magnetic and  kinetic energies in time in particular simulation runs
   of continually driven systems:  \cite{Ng11} (left) and \cite{R07, R08} (right).  
   In both cases time is measured in units of the Alfv{\'e}n 
   crossing time along the loop.  Shown in the figures and common to all  simulations  of Section~\ref{subsec:cont}
   is a intermittent fluctuation of the quantities about average values, with a dominance of magnetic
   energy over kinetic energy. 
  Left: image adapted from \cite{Ng11}.
  Right: image reproduced with permission from \cite{R08}, copyright AAS.
   }
   \label{fig:contmagen}
\end{figure}

Several features are common to each of the systems described in this section. The most fundamental feature is that
after a fairly short time statistically steady states are reached where quantities (e.g. total magnetic and kinetic
energies, dissipation) fluctuate in time about an average level.
The average levels as well as the character of the intermittency both depend on the simulation details.
Figure~\ref{fig:contmagen} shows the volume magnetic and kinetic energies in time for particular runs from \cite{Ng11}
and \cite{R07, R08}.
  In these and other cases magnetic energy dominates significantly over kinetic 
energy (example ratios $10:1$ \cite{LS93}, $40:1$ \cite{Ng11, R07, R08}, $80:1$ \cite{B13}).
Rappazzo {\it et al.} \cite{R07} find the power spectrum of the total energy depends on 
the typical driving velocity compared with the Alfv{\'e}n speed, steepening as
the driving velocity is comparably increased (in common with \cite{GN96}).

Ng \textit{et al.}  \cite{Ng11} show that the average free magnetic energy levels increases with
 magnetic Reynolds number  (Figure~\ref{fig:contmagen}, left).
Noteworthy for comparison is that for the continually applied stationary shearing profile of Rappazzo \textit{et al.} \cite{R10}
a significant amount of magnetic energy initially builds up until a tearing-mode-like instability occurs.  At that point
a sizeable proportion of the free energy is released but further driving leads only to a statistically steady state
and no further significant free energy develops (see Figure 1 of \cite{R10}).  
Hence after the initial instability the system evolution is broadly
similar to those where a complex rotational braiding flow is applied.  A similar finding occurs for the localised single
vortex driver after the initial kink instability \cite{R13}.
This suggests the nature of the photospheric motions are not important for loop heating, in contrast to 
other results \cite{WS11} that will be discussed in Section~\ref{sec:initially}.

 \begin{figure}[htbp]
   \centering
   \includegraphics[width=0.44\textwidth]{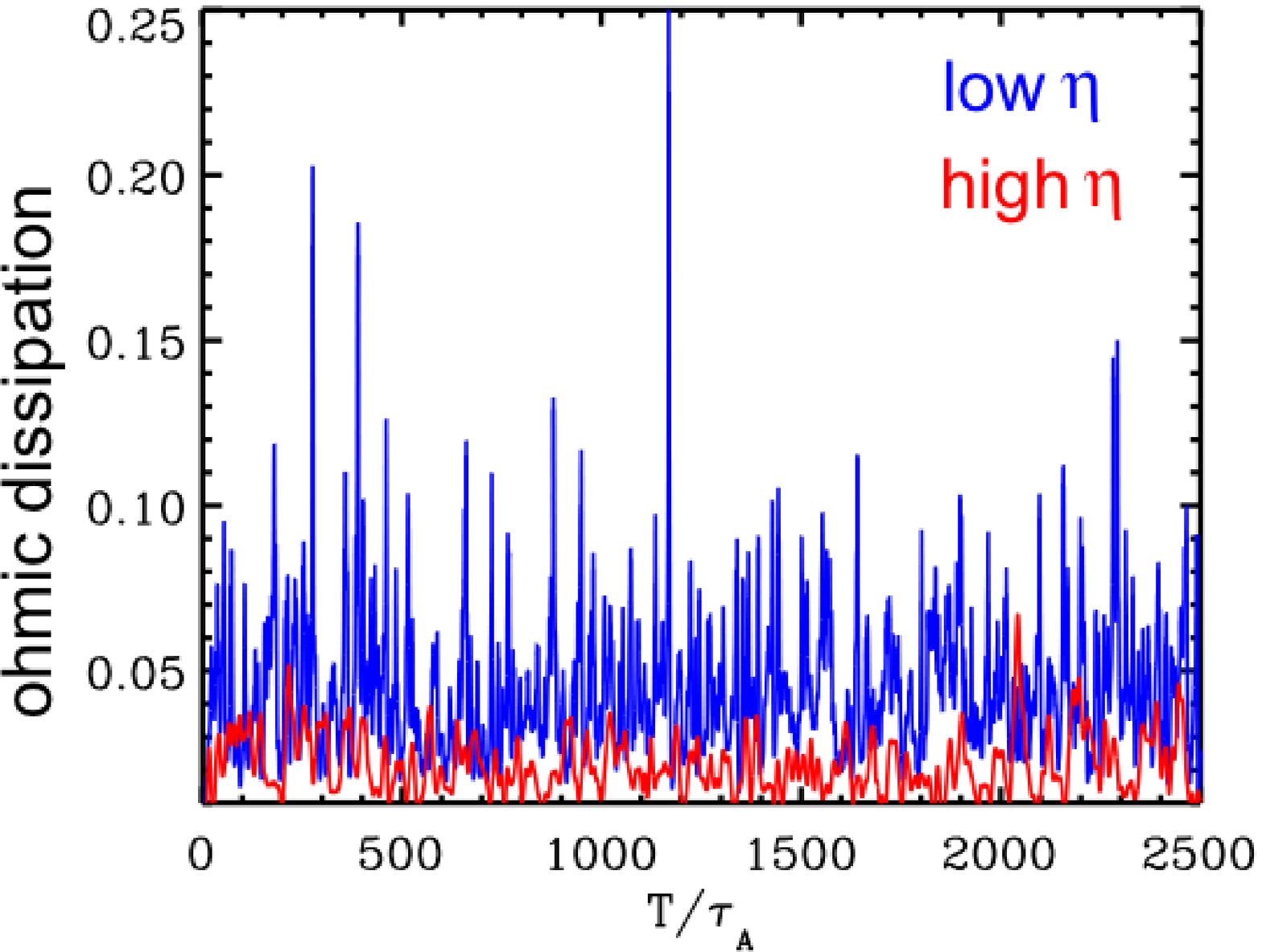}  \ \ \ \ \ 
   \includegraphics[width=0.44\textwidth]{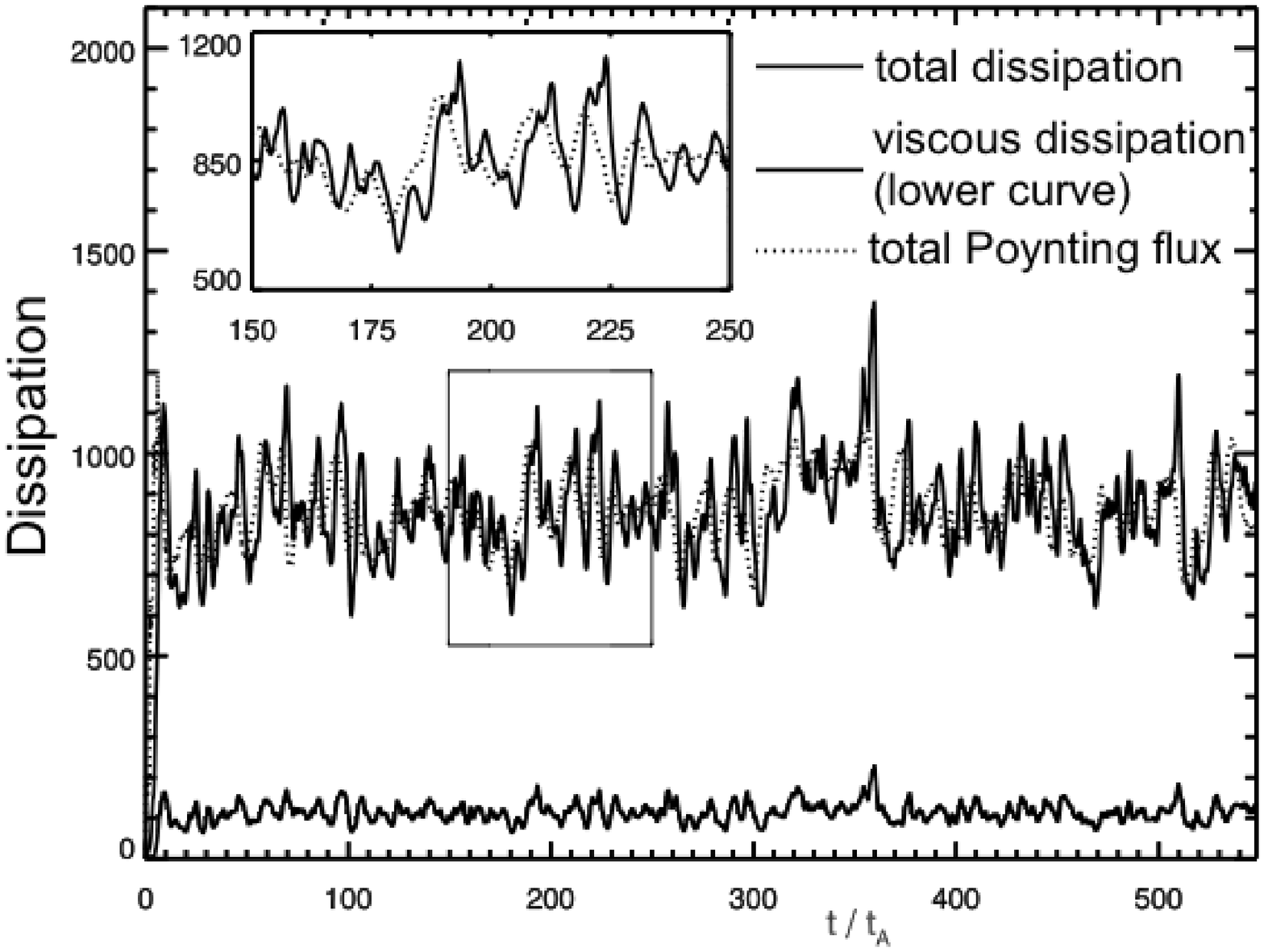} 
   \caption{
Dissipation shows a bursty, intermittent character in continually driven systems.   
Left:  The level of dissipation increases with decreasing magnetic Reynolds number seen in the example of
 \cite{Ng11} (image adapted from \cite{Ng11}).  
Right:   Dissipation and Poynting flux balance only on long timescales but are decoupled
 on short periods, seen in the example of \cite{R07,R08} (image adapted from \cite{R07}). }
   \label{fig:contdiss}
\end{figure}

Dissipation in these systems has a similar character, with both ohmic and viscous dissipation having a bursty,
intermittent nature, illustrated for some example runs in Figure~\ref{fig:contdiss}.  
Poynting flux into the volume and dissipation are coupled only on long time periods, with the
de-correlation being particularly noticeable during strong heating events \cite{R07}.
The average heating rate is found to increase and the fluctuations to become increasingly
fast as the magnetic Reynolds number increases \cite{Ng11} (see also Figure~12 of \cite{R08}, for example).
The dependency on resistivity, $\eta$, was examined systematically first by
Longcope \& Sudan \cite{LS93} and then over a wider range of $\eta$ given the newly available increased 
computing power by Ng \textit{et al.} \cite{Ng11}.  
The Ng \textit{et al.} \cite{Ng11} results show that the $\eta^{-1/3}$ dependency found
by Longcope \& Sudan \cite{LS93} begins to turn over as $\eta$ is decreased.  The
 behaviour over further orders of magnitude in $\eta$ is far from clear but the data clearly
warn against extrapolating the few available points to solar parameters.

 \begin{figure}[htbp]
   \centering
   \includegraphics[width=0.44\textwidth]{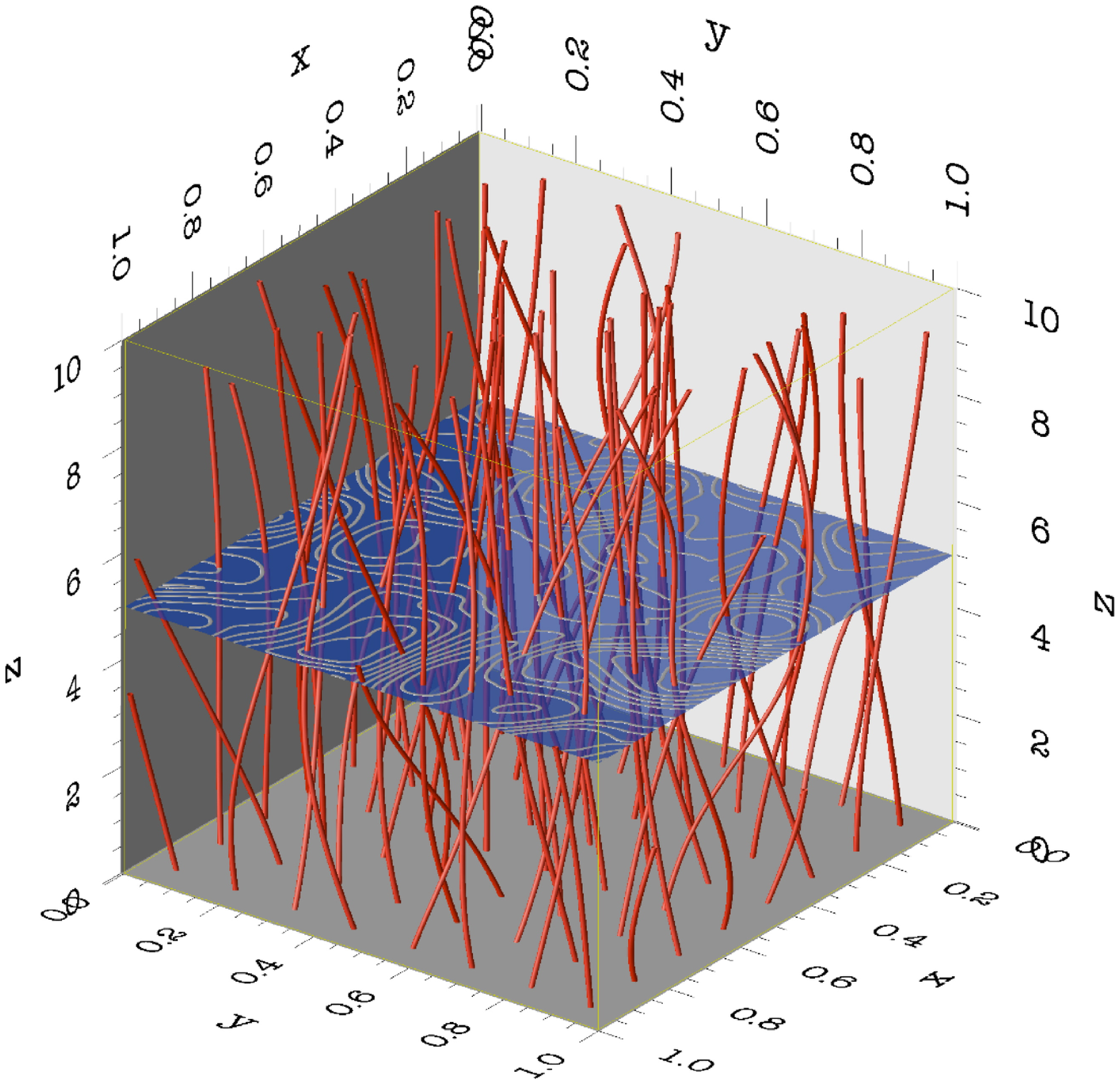} 
      \includegraphics[width=0.44\textwidth]{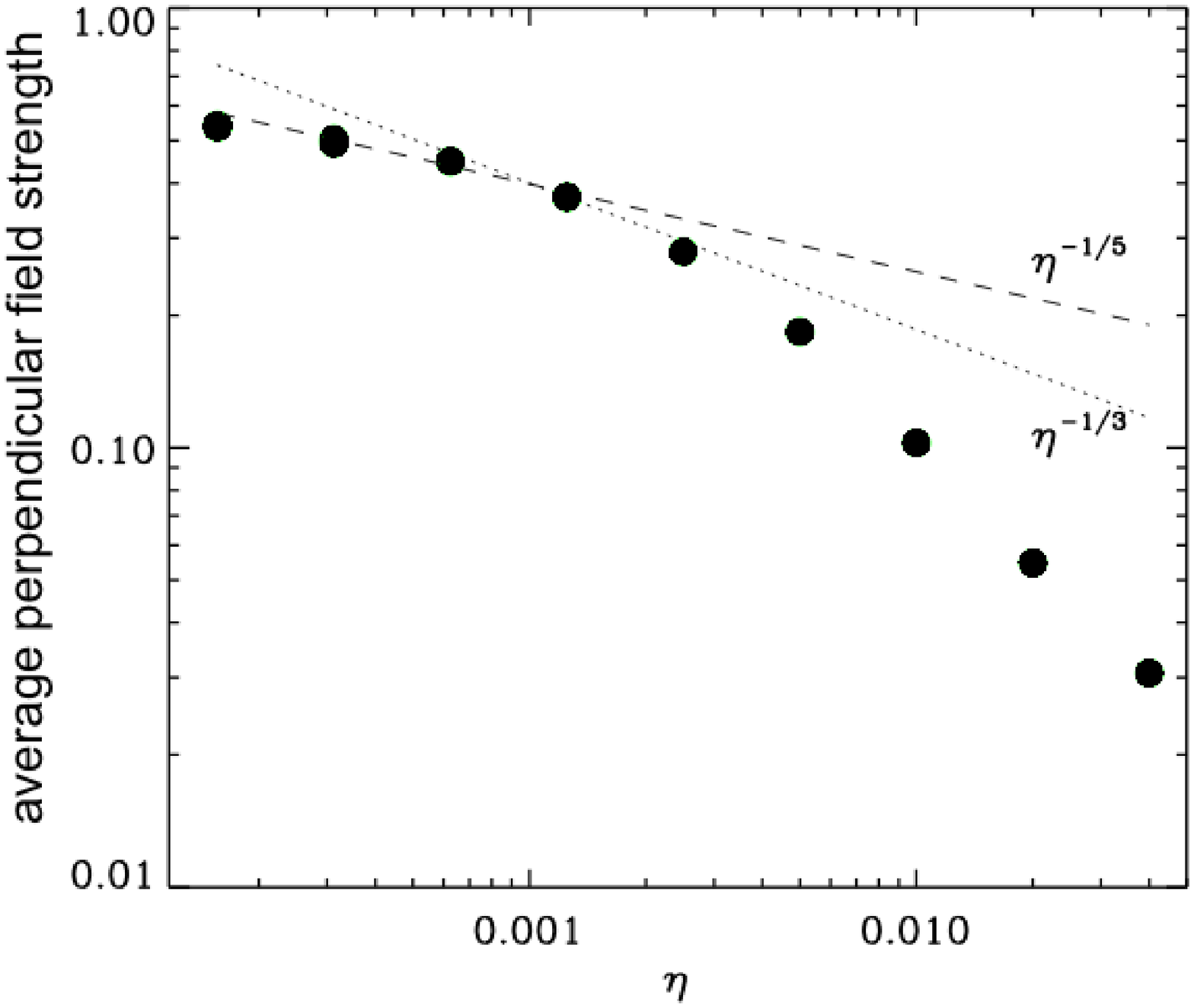} 
   \caption{Left: example magnetic field lines in a continually driven system of  Rappazzo \textit{et al.}
\cite{R07, R08} showing low levels of braiding in the statistically steady state.   Image 
reproduced with permission from \cite{R08}, copyright AAS.
Right: A detailed examination of $B_{\perp}$ in Ng \textit{et al.}'s  reduced MHD simulations \cite{Ng11}
shows an increase of average $B_{\perp}$ as $\eta$ decreases.
Image adapted from \cite{Ng11}.}
   \label{fig:contfield}
\end{figure}

Linked with both the dissipation and the magnetic energy is the magnetic field structure.  Rappazzo \textit{et al.}
\cite{R07, R08} find that the magnetic field stays close to uniform, with typical inclinations being of the order 
$2^{\circ}$, illustrated in Figure~\ref{fig:contfield} (left).  
While \cite{R07, R08} do
not detail how the structure depends on  magnetic Reynolds number,
Ng \textit{et al.} \cite{Ng11} examines this more closely, showing how the average perpendicular field strength $B_{\perp}$
increases with decreasing  $\eta$ (Figure~\ref{fig:contfield}, right).
Note that a basic assumption of RMHD is that $B_{\perp}$ is an order of magnitude less than
$B_{z}$ under the evolution and hence the findings lead one to ask whether this trend would also occur 
in a full 3D MHD evoulution.

The current structure in these continually driven systems \cite{LS93, HH96, G00, R07, R08, Ng11, D12} 
are broadly similar, with long thin current layers extending vertically through much of the domain
(see Figure~6 of \cite{G00}, Figure~18 of \cite{R08}, Figure~1 of \cite{Ng11}, for example). 
  Part of this structure is enforced by the basic RMHD assumption where only the vertical current component is 
  retained. %(the vertical extent of the component is not enforced).  
  In the continually driven shearing simulations 
  of Galsgaard \& Nordlund \cite{GN96} a much  more fragmented and intermittent structure is found (Figure~\ref{fig:GN96} 
  here, see also Figure~9 of \cite{GN96}).
    However, although a fully 3D MHD evolution was considered in \cite{GN96}, the
    initial plasma beta was set at $\beta =0.5$ and the
  energy equation neglected conduction and radiation so that a high--$\beta$ state developed.  
 Note that in the reduced MHD simulations the lack of an energy equation essentially implies $\beta=0$ throughout.
 The only fully 3D MHD long duration simulations including conduction and radiation to date seem to be
those of Dahlburg \textit{et al.}  \cite{D12}.  
These authors are the first to show that the internal energy also has an intermittent,
statistically steady nature.  The temperature of their coronal loops corresponds closely to 
the current structure and is found to be highly spatially structured, giving a multi-thermal loop where only
a fraction of the volume shows significant heating at any one time.   Dahlburg \textit{et al.} \cite{D12} find 
\textit{``the dynamics of this problem are well represented by RMHD''} with low average twist (maximum $3^{\circ}$)
and an almost incompressible flow, although a detailed comparison is not made in the letter.

\subsection{Formation of Discontinuities}

A third class of simulations consider the Parker problem, i.e.~whether or not truly singular currents form
under an ideal MHD evolution where a topologically simple magnetic field is subjected to 
complex boundary motions.  Several of the sequences of boundary shear works 
discussed in Section~\ref{subsec:shear}  address this question.  In summary, 
successive application of low amplitude shear current layer thickness exponentially decreases with 
number of shears (i.e.~finite thickness currents) \cite{vB88a,vB88a,MSvH}.
For sufficiently strong shear behaviour is consistent with the formation of singular current sheets,
up to the numerical resolution available \cite{L98}.

\begin{figure}[htbp]
   \centering
     \includegraphics[width=0.4\textwidth]{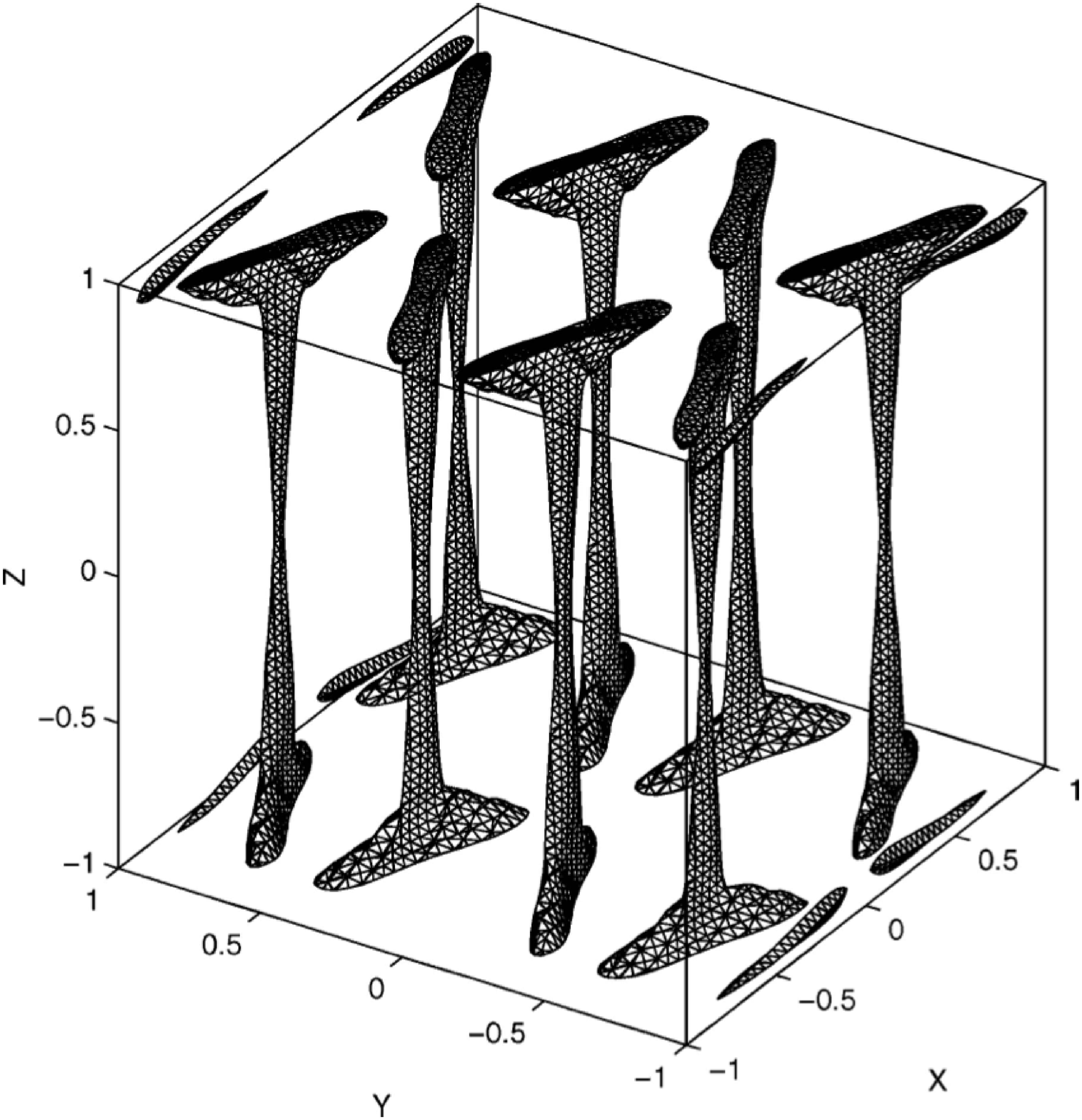} 
     \includegraphics[width=0.45\textwidth]{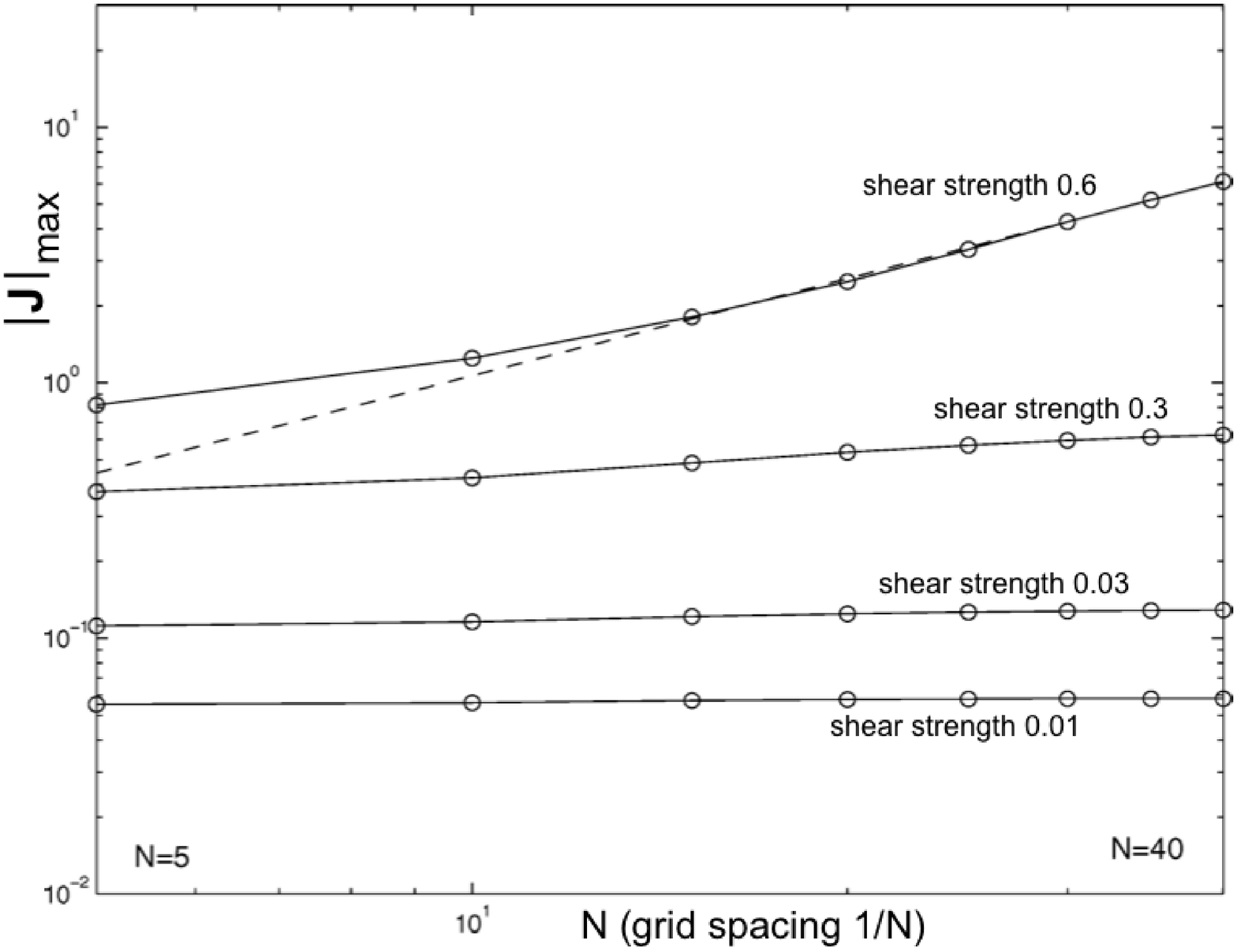} 
   \caption{Left:  Isosurfaces of current density in an equilibrium state following localised boundary motions of shear 
   and compression applied to a uniform field and obtained under a magnetofrictional relaxation.  
   A wide variety of boundary motions considered also resulted in smooth equilibria \cite{CS05}.  
   Right:  Only for footpoint motions also involving the side boundaries was behaviour consistent with 
   singular current sheet formation -- illustrated by the power-law growth for sufficiently strong shear.   
  Images reproduced with permission from \cite{CS05}.}
   \label{fig:CS05}
\end{figure}

Moving to the more general class of simulation, Craig \& Sneyd \cite{CS05} presented a series of simulations
where a wide variety of boundary motions were applied to a uniform magnetic field on a unit cube,
again using the magnetofrictional relaxation technique \cite{CS86, L98}.
In almost all cases considered smooth, well-resolved, large-scale currents are obtained.
For example, following a particular combination of localised shear and compression the current
structure shown in Figure~\ref{fig:CS05} (left) is found.    
%Smooth current fingers run vertically through
%the domain (primarily concentrated at the footpoints of the loop).
Only for one type of footpoint motion do the authors find behaviour that is consistent with the formation of a singular
current sheet, this being where footpoint displacements involve the side-boundaries of the loop.  Under these 
circumstances and given motions of sufficient amplitude, the maximum current in a relaxed equilibrium increases
with grid resolution (to the maximum available resolution of $81^{3}$), as illustrated in 
Figure~\ref{fig:CS05} (right).
This finding is in common with Longbottom \textit{et al.} \cite{L98}. However the authors point out that 
the types of motions employed in both works are not those originally envisaged under the Parker 
problem (i.e.~they are not complex motions internal to a loop but rather involve the entire loop).

With this in mind a series of works by the group in Dundee have examined the nature of currents
in magnetic fields that are braided \cite{WS09a, WS09b, WS10, P10, Y10, WS11, P14+}.
  The idea differs from previous simulations in that, rather than
starting with a uniform magnetic field and braiding it through boundary motions, the initial configuration
is already braided \cite{WS09a} (chosen as an analytical expression where some magnetic field lines have a pigtail braid
topology while more generally field lines show a complex continuum of connectivities).  
 Overall the configuration has no helicity, being built up from an equal number of positive 
and negative twists.  The energy required to create the field is low, with about 3\% magnetic energy in excess of potential \cite{WS09a}.
The field line mapping of the braided magnetic field shows small scales, with the thickness of the 
QSLs in the field decreasing exponentially with braid complexity \cite{WS09b}, mirroring the finding of 
van Ballegooijen \cite{vB88a} for the multiple shear case.
The authors then ask whether the pigtail braid configuration can be ideally deformed to a force-free state:  if so then
the space of force-free fields is surely not as restricted as Parker envisaged.  To carry out this evolution 
the aforementioned magnetofrictional relaxation scheme was employed \cite{WS09a}.
The ideal relaxation is able to bring the braided field to a near force-free state where current 
structures show only large scales (Figure~\ref{fig:BraidJ}, left-most image). 
However, numerical difficulties with the scheme \cite{P09}
prevent relaxation to perfectly force-free (a state that anyway can only be asymptotically 
reached in a simulation).  A more recent
detailed investigation has suggested that as the force-free state is approached even more closely, the 
current structure would develop small scales, with the thickness of current layers having
the same width as the QSLs of the field and so exponentially decreasing with level of braiding \cite{P14+}.  
Again this mirrors the finding of van Ballegooijen \cite{vB88b}. 
In any case, in a resistive MHD simulation the current is found to collapse and reconnection to 
begin across the resulting layers  \cite{WS10}.  The subsequent evolution of the field,
together with other similar simulations,  is the subject of the next  section.

\subsection{Initially braided fields}
\label{sec:initially}

\begin{figure}[htbp]
   \centering
     \includegraphics[width=\textwidth]{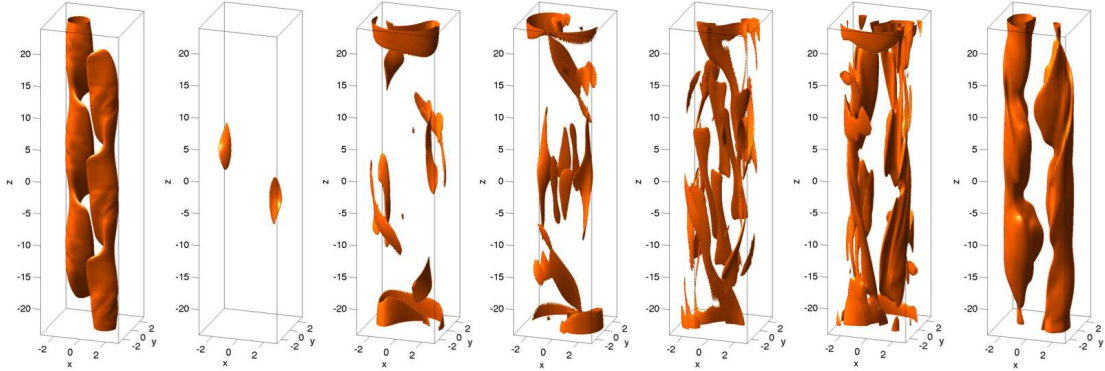} 
   \caption{Isosurfaces of current in time  ($t=0, 15, 27, 35, 50, 140, 290$)  during a resistive MHD relaxation
   of a braided magnetic field.  After \cite{P10}.}
   \label{fig:BraidJ}
\end{figure}

\begin{figure}[htbp]
   \centering
     \includegraphics[width=0.8\textwidth]{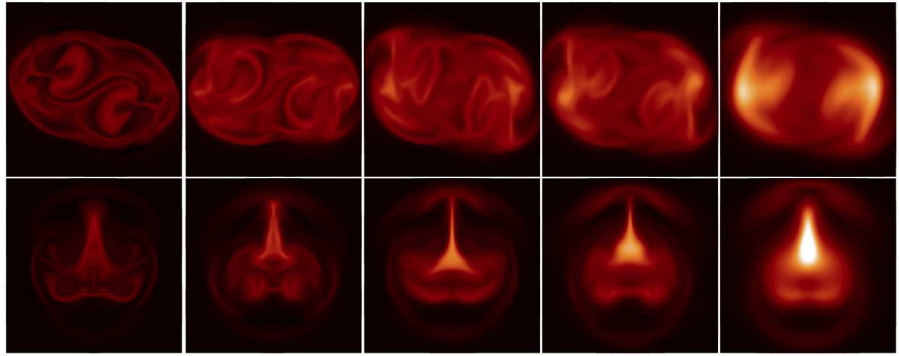} 
   \caption{Loop heating in time ($t=40, 80, 120, 180, 350$) for two contrasting resistive relaxations of braided 
   magnetic fields.  The upper panel shows the homogeoneous heating of a complex braided field and the
   lower panel the more localised intense heating for a more coherently braided field.   
   Field line averaged temperature along the loop is shown in the loop mid-plane. After \cite{WS11}.}
   \label{fig:BraidHeat}
\end{figure}

The near-force-free magnetic field with a pigtail braid topology described in the previous section
has been used as an initial condition for a resistive MHD evolution \cite{WS10, P10}.  
At the magnetic Reynolds numbers presently numerically accessible 
the current system quickly % (within $1/4$ of an Alfv{\'e}n travel time across the loop)
 intensifies and collapses to form two thin current layers \cite{WS10}.  
These layers then fragment and a complex network of current layers with a volume filling effect is formed (Figure~\ref{fig:BraidJ}).  
Magnetic reconnection taking place across these layers allows the 
 magnetic field to simplify and form an equilibrium state consisting of two unlinked flux tubes of opposite twist, associated 
with large scale currents (Figure~\ref{fig:BraidJ}, right-most image) \cite{P10}.  
Thus, although helicity is very well conserved, relaxation is not to the expected Taylor
state,  limiting the magnetic energy release \cite{Y10}.
  The braiding is found to be associated with a  homogeneous loop heating \cite{WS11}. By contrast a
comparison case of a more coherent braided field, constructed from twisting motions of only one sign, was 
examined and found to lead to more localised but stronger heating \cite{WS11}.  These contrasting cases, 
illustrated in Figure~\ref{fig:BraidHeat} suggest that the nature of photospheric motions will indeed have a strong impact 
on heating via magnetic braiding.

\begin{figure}[htbp]
   \centering
     \includegraphics[width=0.47\textwidth]{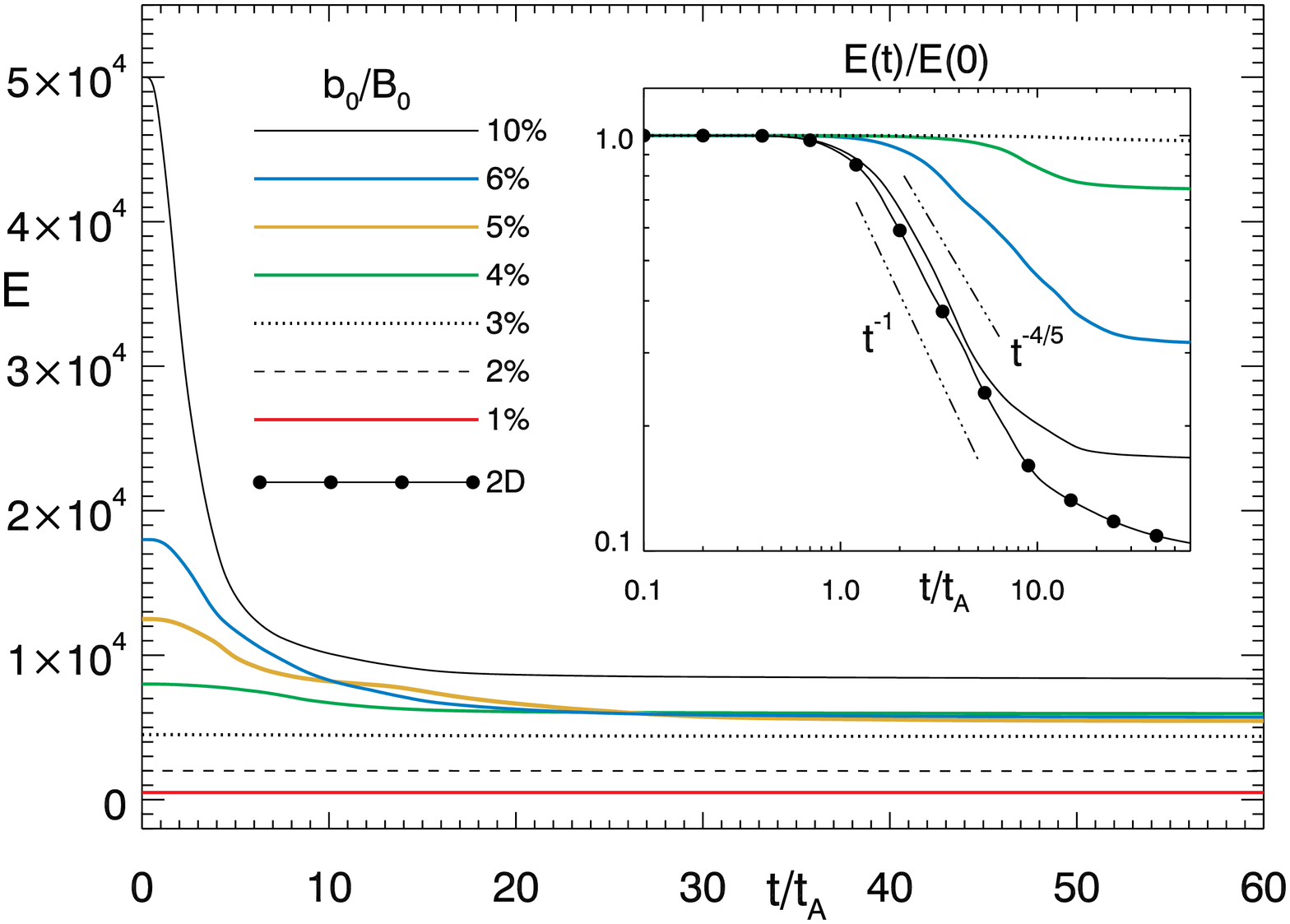} 
          \includegraphics[width=0.49\textwidth]{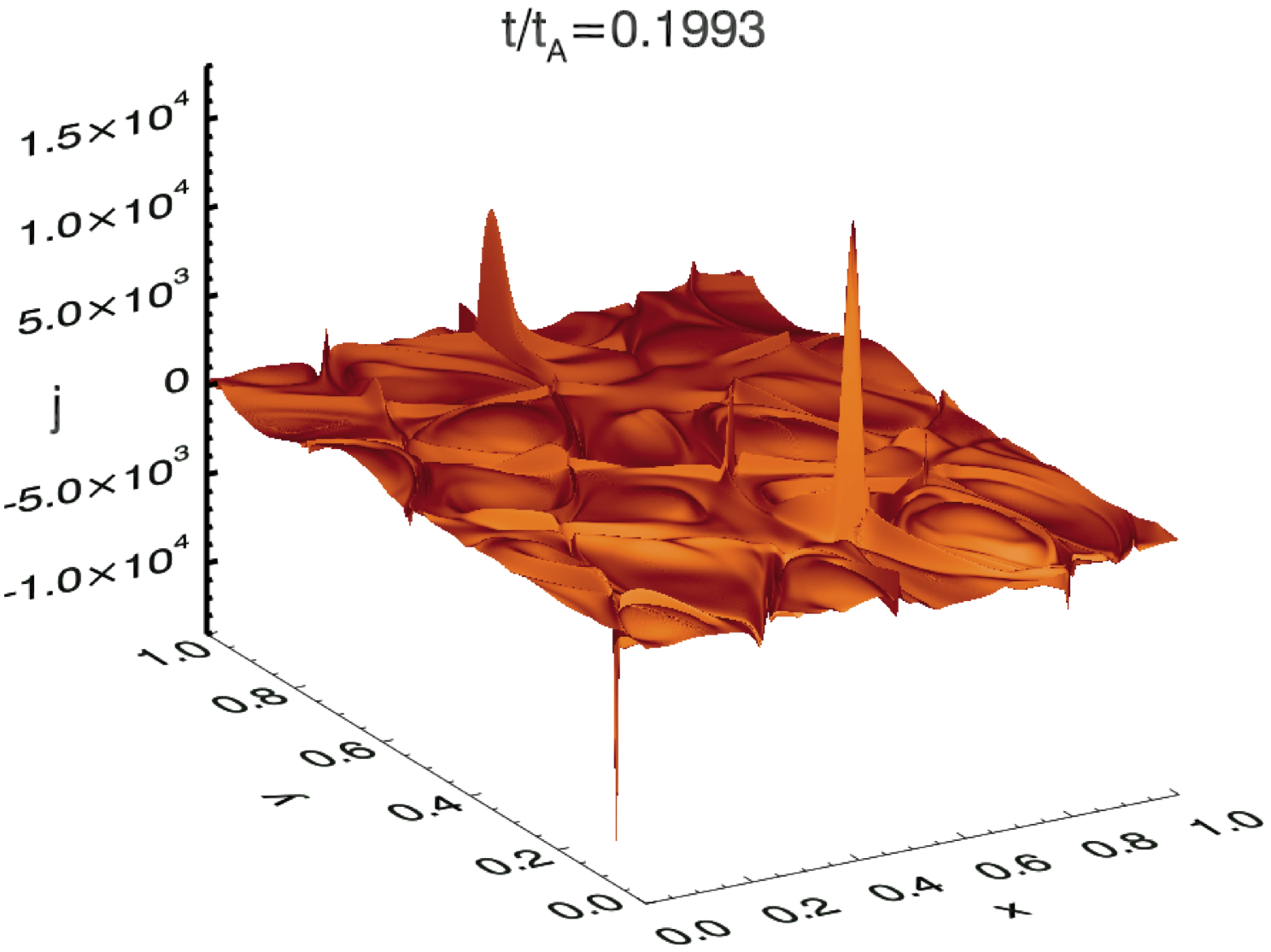} 
   \caption{(Left) Total magnetic energy in time for braided non-equilibrium magnetic fields
   under a resistive reduced MHD evolution.  The ratio $b_{0}/B_{0}$ relates to the initial energy
   of the perpendicular field component.  For sufficiently high $b_{0}/B_{0}$ a turbulent decay
   releases much of the free energy.   
   (Right)  Current $j_{z}$ after $t/t_{A}=0.1993$ in the ideal reduced MHD simulation ($4096^{2}\times2048$ resolution),
   taken in the mid-plane of the loop.  Strong and localised current layers  form but it cannot be established
   whether or not they are singular.
Images reproduced with permission from \cite{RP13}, copyright AAS.}
   \label{fig:RP13}
\end{figure}

In a broadly similar spirit  Rappazzo \& Parker \cite{RP13} present a series of simulations following the resistive reduced
MHD evolution of braided magnetic fields.  The fields, not themselves in equilibrium, are constructed by 
superimposing a perpendicular field component made up of large scale Fourier modes to the background 
uniform field, so creating a braided loop (taken with aspect ratio $1:10$).
The loop evolution was found to depend on the ratio, $b_{0}/B_{0}$, between the root mean square amplitude of the 
initial perpendicular field and the background field.  For $b_{0}/B_{0} \gtrsim 4\%$
the system develops current layers and a resistive turbulent decay ensues, with some but not all of the
free magnetic energy being dissipated (see Figure~\ref{fig:RP13}).
The authors suggested that the ratio $b_{0}/B_{0}$ required before a violent decay could take place
 depends on the loop aspect ratio, as $\sim l/3.5L$ (where $l$ is the horizontal and $L$ the vertical loop dimension).
Next  Rappazzo \& Parker \cite{RP13}  consider whether the current layers forming in the early 
resistive RMHD evolution would be singular in an ideal evolution.  For this the same RMHD scheme was deployed,
taking a high resolution and only numerical dissipation.  The analyticity strip method \cite{SSF83} 
was applied to examine whether the nature of the current evolution is indicative of a singularity, but no 
conclusion could be drawn, with the method itself failing.  Nevertheless the early evolution shows strong and 
highly localised current enhancements, as shown in Figure~\ref{fig:RP13} (right).

\section{Conclusions and Future Directions}
\label{sec:conclusions}

The emerging consensus from flux braiding experiments is that thin but non-singular current
layers form as coronal loops are subjected to braiding motions and that the width of these
layers decreases exponentially in time.  With photospheric motions continually but slowly braiding
the coronal volume dissipation is an inevitable consequence.  
In simulations where systems have a finite resistivity and are subjected to footpoint motions over
a long period of time statistically steady states are reached where dissipation and magnetic energy
fluctuate strongly in time about an average level.  
Although the Poynting flux into the corona must balance over a long time (with short-term decoupling present 
in simulations)  the flux is itself dependent  on the state of the coronal field.  How exactly heating depends on 
the nature of the photospheric motions and on the coronal resistivity is not well understood.

A number of ideas arise for future attacks on the braiding problem.  One is to update
the simulation approach to the singularity question of Parker using improved computational
techniques.   For example, the magnetofrictional relaxation scheme, previously known to have 
numerical inaccuracies, has been significantly improved and at the same time parallelised so that
simulations with an order of magnitude higher resolution are possible \cite{C14}.

For the case where loops are continually subjected to boundary motions a more detailed examination with 
systematic simulation setup is
required to determine what exactly are the important factors in determining the level and nature of the
loop heating.  Before tackling this with a reduced MHD approach a careful comparison of reduced MHD with
fully 3D MHD could perhaps be fruitful.  

In a move towards increased realism a basic question is how the corona can be braided when a realistic 
loop is modelled.  Part of this is to take a representative stratified atmosphere, with important initial advances
recently presented by van Ballegooijen \textit{et al.} \cite{vB14}.
Additionally one should recall that the real corona is full of topological structure so that how coronal fields
are braided when the magnetic carpet is included, i.e.~simulations addressing the coronal tectonics hypothesis \cite{Priest},
 is another task for the near future.
Similarly along these lines the ever-increasing computing power should allow a better resolution of large-scale simulations
(e.g. \cite{GN02, PGN04, GN05, GN05b}) so that their relation to braiding and coronal tectonics can be 
more completely established.

\section*{Acknowledgment}

Helpful discussions with Peter Cargill, Jim Klimchuk, and the Dundee MHD group 
 and financial support from STFC (ST/K000993/1)  are all gratefully acknowledged.

\end{document}